\documentclass[epj]{svjourmod}

\usepackage{amsmath,amssymb,amsfonts,dcolumn,mathcomp,slashed,dsfont}
\usepackage{graphicx}
\usepackage{citesort}
\usepackage{psfrag}
\usepackage{multirow}
\usepackage{booktabs}
\usepackage{bm}

\newcommand{\beq}{\begin{equation}}
\newcommand{\eeq}{\end{equation}}
\newcommand{\M}{\mathcal{M}}
\newcommand{\T}{\mathcal{T}}
\newcommand{\F}{\mathcal{F}}
\newcommand{\G}{\mathcal{G}}
\newcommand{\N}{\mathcal{N}}
\newcommand{\A}{\mathcal{A}}
\newcommand{\BR}{\mathcal{B}}
\newcommand{\Lagr}{\mathcal{L}}
\newcommand{\Fpi}{F_\pi}
\newcommand{\Order}{\mathcal{O}}
\newcommand{\disc}{{\rm disc}\,}
\newcommand{\mpn}{M_{\pi^0}}
\newcommand{\mpc}{M_\pi}
\newcommand{\mV}{M_V}
\newcommand{\mr}{M_{\rho_3}}
\newcommand{\mrr}{M_{\rho'}}
\newcommand{\mrrr}{M_{\rho''}}
\newcommand{\grr}{\Gamma_{\rho'}}
\newcommand{\grrr}{\Gamma_{\rho''}}
\newcommand{\nnnl}{\nonumber\\}
\renewcommand{\Re}{{\rm Re}\,}
\renewcommand{\Im}{{\rm Im}\,}

\def\Xint#1{\mathchoice
   {\XXint\displaystyle\textstyle{#1}}%
   {\XXint\textstyle\scriptstyle{#1}}%
   {\XXint\scriptstyle\scriptscriptstyle{#1}}%
   {\XXint\scriptscriptstyle\scriptscriptstyle{#1}}%
   \!\int}
\def\XXint#1#2#3{{\setbox0=\hbox{$#1{#2#3}{\int}$}
     \vcenter{\hbox{$#2#3$}}\kern-0.5\wd0}}
\newcommand{\dashint}[1]{\Xint{\hspace{#1}-}}

\begin{document}

\title{Dispersive analysis of \boldmath{$\omega\to 3\pi$ and $\phi\to 3\pi$} decays}
\titlerunning{Dispersive analysis of $\omega\to 3\pi$ and $\phi\to 3\pi$ decays}

\author{Franz Niecknig, Bastian Kubis, Sebastian P.\ Schneider}

\institute{
   Helmholtz-Institut f\"ur Strahlen- und Kernphysik (Theorie)
   and 
   Bethe Center for Theoretical Physics,
   Universit\"at Bonn, \linebreak D--53115~Bonn, Germany
}

\authorrunning{F.\ Niecknig, B.\ Kubis and S.\ P.\ Schneider}

\date{
}

\abstract{We study the three-pion decays of the lightest isoscalar vector mesons, $\omega$ and $\phi$, 
in a dispersive framework that allows for a consistent description of final-state interactions 
between all three pions. 
Our results are solely dependent on the phenomenological input for the pion--pion P-wave scattering phase shift. 
We predict the Dalitz plot distributions for both decays and
compare our findings to recent measurements of the $\phi\to3\pi$ Dalitz plot by the KLOE and CMD-2 
collaborations.
Dalitz plot parameters for future precision measurements of $\omega\to3\pi$ are predicted.
We also calculate the $\pi\pi$ P-wave inelasticity contribution from $\omega\pi$ intermediate states. 
\PACS{
      {11.55.Fv}{Dispersion relations}
      \and
      {13.25.Jx}{Hadronic decays of other mesons}
      \and
      {13.75.Lb}{Meson--meson interactions}
     }
}

\maketitle

\section{Introduction}

The treatment of meson decays into three pions using dispersion relations is a classic subject.
It was developed first in the 1960s in the context of $K\to 3\pi$ decays~\cite{KhuriTreiman},
and already in the 1970s, it was applied specifically to the decay $\omega\to 3\pi$~\cite{AitchisonGolding}.
One of its main virtues is the fact that final-state interactions among the three pions
are fully taken into account, in contrast to perturbative, field-theory-based
approaches; the constraints coming from analyticity and unitarity are respected exactly to all orders.
This becomes the more important, the higher the mass of the decaying particle, hence the higher
the possible energies of the two-pion subsystems within the Dalitz plot:
while decays like $K\to 3\pi$ or $\eta\to 3\pi$ have successfully been analyzed
in perturbative settings like chiral perturbation theory~\cite{GLeta3pi,BGeta3pi}
or even non-relativistic effective theories~\cite{CGKR,GKR,etaDal}, it is obvious that 
these are doomed to fail for the decays of $\omega$ or $\phi$ into three pions, where
the influence of the $\rho$ resonance is already strongly felt ($\omega$) or even 
shows up in the form of resonance bands inside the Dalitz plot ($\phi$).

In recent years, there has been a flourish of attempts to also treat the physics of vector mesons 
in effective-field-theory approaches~\cite{Bijnens1,Bijnens2,Bijnens3,Bruns,Lutz,Leupold,Mainz,Kampf}, motivated not least
because of their prominent nature in hadronic physics involving virtual photons, 
their relatively low masses among the various meson resonances,
and their strong coupling to the light pseudoscalars.
However, when it comes to the three-pion decays of $\omega$ and $\phi$, 
descriptions in terms of vector-meson dominance and improved tree-level models
(mostly incorporating a finite width of the intermediate $\rho$ resonances)~\cite{Leupold,Klingl,Benayoun1,Benayoun2}
(see also Ref.~\cite{UGM-PhysRept} for earlier references),
which do not fully respect analyticity and unitarity constraints,
still seem to be state of the art.\footnote{It needs to be pointed out, though, that Lagrangian-based approaches
often have the advantage of relating various different processes to each other; such symmetry constraints most 
of the time have nothing to do with analyticity and unitarity, and hence can at best be imposed \emph{a posteriori}
in dispersive studies.}

\begin{sloppypar}
In this respect, we consider this an ideal time to take up dispersive studies of these decays once more.
High-precision phase-shift analyses of pion--pion scattering are now available~\cite{ACGL,CapriniWIP,Pelaez}
and can be employed as input for decay studies.  On the experimental side, 
high-statistics Dalitz plot investigations have either been performed  
($\phi\to 3\pi$~\cite{KLOE,CMD-2}), or are planned or ongoing ($\omega\to 3\pi$~\cite{WASA,KLOE2}).
Finally, these systems are ideal test cases to study the achievable precision of Dalitz plot descriptions
with theoretically rigorous methods: they are simple in terms of partial waves, 
as entirely P-wave dominated, and kinematically in the transition regime between the light (kaon, eta) 
decays and those of heavy ($D$, $B$) mesons that might be treated with similar methods in the future~\cite{Hadron11}.
\end{sloppypar}

The outline of this article is as follows. 
We introduce the necessary basics on kinematics and partial-wave decomposition in Sect.~\ref{sec:V3pikin},
and describe the formalism for a dispersion-theoretical description of the decays $\omega,\,\phi\to 3\pi$ 
in Sect.~\ref{sec:Disp3body}.
First numerical results of our solutions are discussed in Sect.~\ref{sec:numres},
before we compare in detail to the experimental Dalitz plot studies for $\phi\to 3\pi$ and predict
Dalitz plot parameters for $\omega\to 3\pi$ in Sect.~\ref{sec:expcomp}.
As a further application, we calculate the contribution of $\omega\pi$ intermediate states
to the inelasticity in the pion--pion P partial wave in Sect.~\ref{sec:inel}.  
We conclude in Sect.~\ref{sec:summary}.  
Several aspects that lie somewhat outside the main line of argument are relegated to the Appendices, 
where we discuss higher partial waves (Appendix~\ref{app:Fwave}),
the analytic properties of the dispersive integral (Appendix~\ref{app:angint}), 
as well as possible contributions from heavier resonances (Appendix~\ref{app:highres}),
and derive a generalization of the Omn\`es representation taking into account inelasticities (Appendix~\ref{app:inpara}).

Preliminary results of this study have already been presented in Ref.~\cite{PhiPsiProcs}.

\section{Kinematics and partial-wave decomposition}\label{sec:V3pikin}

In this work we consider the three-pion decay of the lightest isoscalar vector mesons, 
\begin{equation}
 V(p_V)\to\pi^+(p_+)\pi^-(p_-)\pi^0(p_0)~,\qquad V=\omega/\phi~,
\end{equation}
where the particle momenta are conventionally related to the Mandelstam variables by 
 $s=(p_V-p_0)^2$, $t=(p_V-p_+)^2$, $u=(p_V-p_-)^2$, with 
\begin{align}
 3s_0\doteq s+t+u=\mV^2+3\,\mpc^2~.
\end{align}
Here and in the following we restrict ourselves to the isospin limit, $\mpn=M_{\pi^\pm}\doteq\mpc$,
unless explicitly stated differently.

Since $V\to3\pi$ is of odd intrinsic parity, the amplitude can be decomposed as
\begin{equation}\label{eq:AmpMF}
 \M(s,t,u)=i\epsilon_{\mu\nu\alpha\beta}n^\mu p_+^\nu p_-^\alpha p_0^\beta \F(s,t,u)~,
\end{equation}
where $n^\mu$ is the polarization vector of the decaying vector particle, and $\F(s,t,u)$ is a scalar function. 
The absolute value squared of the amplitude reads
\begin{align}
 |\M(s,t,u)|^2 &=\frac{1}{4}\Big[ s\,t\,u-\mpc^2\big(M_V^2-\mpc^2\big)^2\Big]|\F(s,t,u)|^2 \nnnl
&=\frac{s}{16}\kappa^2(s)\sin^2\theta_s|\F(s,t,u)|^2~, \label{eq:phasesp}
\end{align}
where $\theta_s$ is the center-of-mass scattering angle in the $s$-channel, 
$\cos\theta_s=(t-u)/\kappa(s)$, and 
\begin{equation}
 \kappa(s)=\sigma_\pi(s)\lambda^{1/2}(\mV^2,\mpc^2,s)~,
\end{equation}
with the K\"all\'en function $\lambda(x,y,z)=x^2+y^2+z^2-2(xy+yz+xz)$ and $\sigma_\pi(s)=\sqrt{1-4\mpc^2/s}$. 

Due to Bose symmetry only partial waves of odd angular momentum can contribute to the amplitude.
Accordingly, the partial-wave decomposition of the scalar part of the amplitude reads~\cite{JacobWick}
\begin{equation}\label{eq:PWEomega}
 \F(s,t,u)=\sum_{\ell\;{\rm odd}}f_\ell(s)P'_\ell(z_s)~,
\end{equation}
where $z_s = \cos\theta_s$, $P'_\ell(z_s)$ is the differentiated Legendre polynomial, 
and one can project onto the partial-wave amplitude $f_\ell(s)$ by
\begin{equation}
 f_\ell(s)=\frac{1}{2}\int_{-1}^{1} d z_s \big[P_{\ell-1}(z_s)-P_{\ell+1}(z_s)\big]\, \F(s,t,u)~.
\end{equation}
For the dominant $\ell=1$ partial wave we have
\begin{equation}
 f_1(s)=\frac{3}{4}\int_{-1}^{1} d z_s \big(1-z_s^2\big)\, \F(s,t,u)~.
\end{equation}
In our analysis we neglect discontinuities of F- and higher partial waves (see the corresponding discussion in Appendix~\ref{app:Fwave}), 
so that $\F(s,t,u)$ can be decomposed in terms of functions of a single variable as
\begin{equation}\label{eq:SVAadd}
 \F(s,t,u)=\F(s)+\F(t)+\F(u)~,
\end{equation}
where $\F(s)$ only has a right-hand cut.
The dispersion relations derived in the following will be given in terms of these single-variable functions $\F(s)$.
Equation~\eqref{eq:SVAadd} represents a symmetrized partial-wave expansion, its generalized form allowing for 
F-wave discontinuities is shown in Appendix~\ref{app:Fwave}, Eq.~\eqref{eq:defP+Famp}.\footnote{Similar amplitude decompositions,
often quoted as being due to a ``reconstruction theorem'', have been demonstrated in the perturbative context of chiral perturbation
theory to two loops for pion--pion scattering~\cite{Stern,Knecht}, and subsequently been extended to $\eta\to3\pi$ 
decays~\cite{AnisovichLeutwyler}.}

We will also make use of the partial-wave expansion of the $\pi\pi\to\pi\pi$ scattering amplitude, 
which is conventionally given as
\begin{equation}\label{eq:PWEpi}
 \T^I(s,t,u)=32\pi\sum_{\ell=0}^{\infty}(2\ell+1)t_\ell^I(s)P_\ell(z)~,
\end{equation}
where $t_\ell^I$ is the partial-wave amplitude of isospin $I$ and angular momentum $\ell$ 
and can be expressed in terms of the phase shift $\delta_\ell^I$ according to
\begin{equation}
 t_\ell^I(s)=\frac{e^{2i\delta_\ell^I(s)}-1}{2i\sigma_\pi(s)}~.
\end{equation}

\section{Dispersive analysis of three-body decays}\label{sec:Disp3body}

In the following, we will present our framework to account for three-body effects in a non-perturbative fashion. 
It relies on the formalism laid out in the context of dispersive studies of $\eta\to3\pi$ 
using the same method~\cite{AnisovichLeutwyler,Lanz}.

Setting up dispersion relations for three-body decays is a non-trivial task due to the very nature of the process in question: 
three-body final states necessitate the incorporation of three-particle cuts, 
a substantial analytic ingredient that is left out in isobaric models, 
like vector-meson dominance models~\cite{Klingl,Leupold}. 
The main idea behind the approach presented here is to set up a set of dispersion relations for the corresponding \emph{scattering} process, i.e.\ $V\pi\to\pi\pi$, with $\mV < 3\mpc$ and $s>(\mV+\mpc)^2$, and 
analytically continue the resulting expressions to the physical realm of the \emph{decay} process, 
$\mV>3\mpc$ and $4\mpc^2\leq s\leq (\mV-\mpc)^2$.

\begin{figure}
\centering
\includegraphics[width=\linewidth]{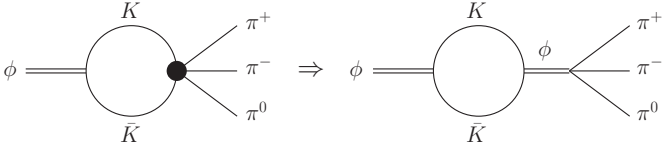}
\caption{Contribution of $K\bar K$ intermediate states to $\phi\to 3\pi$.  
The general rescattering $K\bar K\to \pi^+\pi^-\pi^0$, denoted by the thick dot in the left diagram, is entirely dominated
by the $\phi$ resonance for $(p_K+p_{\bar K})^2 = M_\phi^2$, see the right diagram.}
\label{fig:KKbar}
\end{figure}
\begin{sloppypar}
We wish to remark that we essentially only consider \emph{elastic} unitarity in the following
(see Sect.~\ref{subsec:error} for an attempt to partially account for inelasticity effects in the 
pion--pion P partial wave). 
One might suspect that the dominant decay $\phi\to K\bar K$ with subsequent inelastic rescattering $K\bar K\to3\pi$, 
see Fig.~\ref{fig:KKbar},
may have a significant impact on the Dalitz plot distribution for $\phi\to 3\pi$. 
This is not the case for the following reason: the $K \bar K$ intermediate state
occurs at a fixed total energy of $(p_K+p_{\bar K})^2 = M_\phi^2$, where the rescattering
$K\bar K\to3\pi$ will be entirely dominated by the (very narrow) $\phi$ resonance.  The diagram
therefore factorizes, and the whole effect of the $K\bar K$ intermediate state can be absorbed
into a complex wave-function renormalization constant for the $\phi$ field, which amounts to 
an unobservable overall (constant) phase of the amplitude.  The Dalitz plot distribution of $\phi\to 3\pi$ remains
therefore entirely unaffected.
[A similar argument was already put forward to disregard inelastic rescattering contributions in 
$\eta'\to\eta\pi\pi$ in Ref.~\cite{etaprime}.]
\end{sloppypar}

Note finally that in the sense of a dispersion-theory analysis, there is no $\rho\pi$ two-body
intermediate state contributing to the $\phi$ decay in a similar fashion as $K\bar K$
depicted in Fig.~\ref{fig:KKbar}: $\rho\pi$ is no distinct state from $3\pi$, the effect 
conventionally encoded that way shows up in our analysis as the resonant two-body
$\pi\pi$ P-wave interaction.  From the point of view of dispersion theory, there is no 
meaningful way to differentiate between $\rho\pi$ and $3\pi$ final states in $\phi$ decays.

\subsection{Unitarity relation}\label{subsec:Unitarity}

\begin{figure}
 \centering
 \psfrag{q}[l][t][1]{\hspace{-2mm}  $q-l$}
 \psfrag{l}[c][c][1]{ $l$}
 \psfrag{pm}[c][c][1]{\hspace{-5mm}  $p_-$}
 \psfrag{pp}[c][c][1]{\hspace{-5mm}  $p_+$}
 \psfrag{pn}[c][c][1]{ $p_0$}
 \psfrag{pV}[c][c][1]{ $p_V$}
\includegraphics[width= 0.6\linewidth]{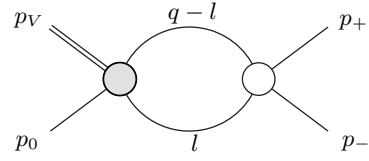} \\
\caption{Diagram of the $V\pi\to\pi\pi$ $s$-channel scattering
amplitude. The double line denotes the vector meson, all single lines refer to pions.
The gray circle corresponds to the $V\pi\to\pi\pi$
amplitude $\M(s,z_s'')$, the white one to the $\pi\pi\to\pi\pi$
amplitude $\T^1(s,z_s')$ (of isospin $I=1$). To calculate the amplitude one cuts through the
intermediate pion propagators with momenta $l$ and $q-l$.}
\label{fig:disc}
\end{figure}

In the following, we relate the discontinuity of the $V\pi\to\pi\pi$ $s$-channel partial-wave amplitude to the amplitude itself and the $\pi\pi$ P-wave phase shift. Similar considerations for $t$- and 
$u$-channel proceed completely analogously. 
In the elastic approximation with only pion--pion intermediate states one has for the discontinuity of the diagram in Fig.~\ref{fig:disc}~\cite{Cutkosky}
\begin{align}\label{eq:optthm}
\disc\M(s,z_s)&=\frac{i}{2}\int\frac{d^4l}{(2\pi)^4}\M(s,z_s') \T^{1*}(s,z_s'') \nnnl
&\times(2\pi)\delta\big(l^2-\mpc^2\big)\,(2\pi)\delta\big((q-l)^2-\mpc^2\big)~,
\end{align}
where $q=p_++p_-$, $z_s=\cos\theta_s$, $z_s'=\cos\theta_s'$, and $z_s''=\cos\theta_s''$, 
where $\theta_s$ denotes the center-of-mass scattering angle between the initial- and final-state momenta, 
$\theta_s'$ between initial and intermediate state, and $\theta_s''$ between intermediate and final
state. 
We remark that in the case of a three-body decay (as opposed to elastic two-body scattering), 
the discontinuity across the cut is in general \emph{not} purely imaginary, so that the usual relation 
$\disc\M(s)=2i\,\Im\M(s)$ does not hold. 
To avoid confusion, we will exclusively refer to the discontinuity of the amplitude.

Upon insertion of Eq.~\eqref{eq:AmpMF} and elimination of the $\delta$-functions,
we can rewrite Eq.~\eqref{eq:optthm} according to 
\begin{align}
\disc\F(s,z_s)&=\frac{i\sigma_\pi(s)}{64\pi^2}\int_{-1}^1 dz_s' \int_0^{2\pi} d\phi_s' 
\frac{z_s''-z_s z_s'}{1-z_s^2} \nnnl &\qquad\qquad \times
\F(s,z_s')\T^{1*}(s,z_s'') ~,
\end{align}
where $z_s''=z_s z_s'+\sqrt{1-z_s^2}\sqrt{1-\smash{z_s'}^2}\cos\phi_s'$. 
Finally, we can use the partial-wave expansion for the amplitudes $\F$ and $\T^1$ 
given in Eqs.~\eqref{eq:PWEomega} and~\eqref{eq:PWEpi}, 
and project out the $\ell=1$ partial wave to arrive at the Watson-like 
unitarity relation,
\begin{equation}\label{eq:Watson}
 \disc f_1(s) = 2i\,f_1(s)\theta(s-4\mpc^2)\sin\delta(s)e^{-i\delta(s)}~,
\end{equation}
where $\delta(s)\doteq\delta_1^1(s)$ is the $\pi\pi$ P-wave phase shift. 
Noting that $\disc f_1(s)=\disc\F(s)$, the previous expression can be recast into
\begin{equation}\label{eq:unrel}
 \disc\F(s) = 2i\,\bigl(\F(s)+\hat\F(s)\bigr)\theta(s-4\mpc^2)\sin\delta(s)e^{-i\delta(s)}~,
\end{equation}
where $\hat\F(s)$ is referred to as the \emph{inhomogeneity} of the integral equation and is given by angular averages over $\F$
according to
\begin{align}\label{eq:inhomogen}
 \hat\F(s)&=3\langle(1-z^2)\F\rangle(s)~,\nnnl
\langle z^n f\rangle(s)&=\frac{1}{2}\int_{-1}^1dz\, z^n f\Bigl(\tfrac{1}{2}(3s_0-s+z\kappa(s))\Bigr)~.
\end{align}
$\hat\F(s)$ contains the left-hand-cut contribution to the partial wave $f_1(s)$.

The previous expressions are well-defined in the scattering regime. However, special care has to be taken in the decay region due to the non-trivial behavior of the function $\kappa(s)$. 
It imposes a complex analytic structure on the angular integration, which is discussed in more detail in Appendix~\ref{app:angint}. 
Aside from that Eqs.~\eqref{eq:unrel} and \eqref{eq:inhomogen} form an integral equation,
the solution of which we discuss in the following.

\subsection{Homogeneous equation, pion vector form factor}\label{subsec:FF}

Before discussing the solution of Eq.~\eqref{eq:unrel} in its entirety, let us consider an obvious simplification, the homogeneous problem with $\hat\F(s)=0$. The unitarity relation then closely
resembles that of the pion vector form factor $F^V_\pi(s)$, and Eq.~\eqref{eq:Watson} is a manifestation of Watson's final-state theorem~\cite{Watson}: the form factor shares the phase of the (elastic) 
scattering amplitude. The solution to this problem is given in terms of the Omn\`es function~\cite{Omnes},
\begin{align}\label{eq:Omnesint}
F^V_\pi(s)&=P(s)\Omega(s)~,\nnnl
\Omega(s)&=\exp\biggl\{\frac{s}{\pi}\int_{4\mpc^2}^\infty ds'\frac{\delta(s')}{s'(s'-s)}\biggr\}~,
\end{align}
where $P(s)$ is a polynomial and the Omn\`es function is normalized to $\Omega(0)=1$. From Eq.~\eqref{eq:Omnesint} we observe that the Omn\`es function is completely 
determined by the phase shift.
In particular, the asymptotic behavior of the Omn\`es function is constrained by the one of the phase shift: 
for $\delta(s\to\infty)\to k\pi$, one finds $\Omega(s)\to s^{-k}$. We shall assume $k=1$ for the asymptotic
behavior of the P~wave, which guarantees the high-energy fall-off for the form factor $\propto 1/s$ as suggested by perturbative 
QCD (up to logarithmic corrections) exactly if $P(s)$ is a constant, and hence due to gauge invariance, $P(s)=1$.\footnote{Note that for a precision analysis of the pion vector form factor up to about $1$\,GeV one needs to account for $\rho$--$\omega$ mixing and the onset of inelasticities. This is not of interest in the context of our work and we refer to Refs.~\cite{Troconiz,Bern:piFF1,Bern:piFF2} for further details.}

Pertaining to this issue the behavior of the $\pi\pi$ phase shift 
is not known to arbitrarily high energies. In our numerical calculations we use a phenomenological parameterization of the phase shift up to a certain energy $s=\Lambda_\Omega^2$, 
beyond which it is set to a constant. At the accuracy at which we are working this constant does not have to be $\pi$ exactly. This issue is studied in Sect.~\ref{subsec:error}.

One possibility to suppress the high-energy behavior of the phase shift in the Omn\`es function is to resort to a twice-subtracted dispersion integral, as opposed to the once-subtracted version
used in Eq.~\eqref{eq:Omnesint} (cf.\ e.g.\ Ref.~\cite{quarkmassFF}). 
Due to the identification $F_\pi^V(s) = \Omega(s)$, the additional subtraction constant can be related to the charge radius of the pion,
\begin{equation}
\Omega(s)=\exp\biggl\{\frac{1}{6}\langle r^2\rangle^V_\pi\, s + \frac{s^2}{\pi}\int_{4\mpc^2}^{\infty}ds'\frac{\delta(s')}{s'^2(s'-s)}\biggr\}~.
\end{equation}
Comparing to Eq.~\eqref{eq:Omnesint}, we can express the charge radius in terms of a sum rule,
\begin{equation}\label{eq:sumrule}
 \langle r^2 \rangle^V_\pi = \frac{6}{\pi}\int_{4\mpc^2}^{\infty}ds'\frac{\delta(s')}{s'^2}~.
\end{equation}
To take advantage of the suppression of high energies in the oversubtracted dispersion integral, 
one may make use of an independent phenomenological determination of the charge radius. 
This allows us to reparameterize our lack of knowledge of the behavior of the amplitude at large $s$ as a polynomial, 
which should be a decent approximation at the energies we are working at. 
In Sect.~\ref{subsec:error} we discuss the numerical effects
of using the two different versions of the Omn\`es function.
We want to emphasize that, as a matter of principle, using a radius different from the sum-rule value Eq.~\eqref{eq:sumrule}
is inconsistent: it leads to a wrong (exponential) high-energy behavior.  In practice, however, and for the (small)
deviations in the charge radius we consider, the effects of this error are not visible within the physical region
of the decays under consideration.

\subsection{Integral equation and solution strategy}\label{subsec:Inteq}

To find a unique solution to the unitarity relation Eq.~\eqref{eq:unrel}, one derives an integral equation for $\F(s)/\Omega(s)$ instead of $\F(s)$ (the issue of ambiguous solutions to the unitarity 
relation is sketched in Refs.~\cite{AnisovichLeutwyler,StefanDiss}). One finds
\begin{equation}
 \disc{\frac{\F(s)}{\Omega(s)}}=\frac{\sin\delta(s)\hat\F(s)}{|\Omega(s)|}\theta(s-4\mpc^2)~,
\end{equation}
the solution of which is readily obtained using Cauchy's integral formula,
\begin{equation}\label{eq:inteeqn}
\F(s)=\Omega(s)\biggr\{a + \frac{s}{\pi}\int_{4\mpc^2}^{\infty}\frac{ds'}{s'}\frac{\sin\delta(s')\hat\F(s')}{|\Omega(s')|(s'-s)}\biggr\}~,
\end{equation}
where $a$ is a subtraction constant. The order of the subtraction polynomial is limited by the asymptotic behavior of the integrand. 
The Froissart bound~\cite{Froissart} constrains the behavior of the amplitude for large $s$,
$\M(s,t,u) < C s \log^2(s)$ for some constant $C$, and consequently $\hat\F(s)<C's^{-1/2}\log^2(s)$ for $s\to\infty$
(and another constant $C'$). Along with the asymptotic behavior of the Omn\`es function it is obvious that the integral remains
finite. We observe that since $\hat\F$ is linear in $\F$ we can further simplify Eq.~\eqref{eq:inteeqn} from the point of view of the numerical implementation:
\begin{align}
\F(s)&=a\F_a(s) ~, \quad \hat\F(s) =a \hat\F_a(s) ~, \nnnl
\F_a(s) &= \Omega(s)\biggr\{1 + \frac{s}{\pi}\int_{4\mpc^2}^{\infty}\frac{ds'}{s'}\frac{\sin\delta(s')\hat\F_a(s')}{|\Omega(s')|(s'-s)}\biggr\}~.
\label{eq:inteeq}
\end{align}
This observation is extremely useful, since our results will be pure predictions aside from an overall normalization constant that can be fixed
\emph{after} the iteration process. 
For lack of a better theoretical method to fix the normalization, we fit to the experimentally determined partial decay width unless explicitly stated otherwise.

The integral equation~\eqref{eq:inteeq} can now be solved by an iterative numerical procedure: we start from an arbitrary input function $\F(s)$ and calculate the inhomogeneity $\hat\F(s)$ from 
Eq.~\eqref{eq:inhomogen}, which in turn is used as an input for the calculation of an updated $\F(s)$ by means of Eq.~\eqref{eq:inteeq}. The process is repeated until the solution converges to a fixed
point with sufficient accuracy. The final result is independent of the specific choice of the starting function. 
We will use $\F(s)=\Omega(s)$ as our starting point: 
as the $\pi\pi$ P~wave is dominated by the $\rho$ resonance at low energies, 
this closely corresponds to an isobaric description of the decay, and 
the modification of $\F(s)$ in the iteration procedure, the difference 
between starting and fixed point, allows us to quantify crossed-channel 
effects generated by the iteration in a plausible way.

\subsection{Oversubtraction}\label{sec:oversubtract}

The integral in the solution of the dispersion relation for $V\to 3\pi$ decays, Eq.~\eqref{eq:inteeq}, 
is guaranteed to converge, given our assumptions on the high-energy behavior of amplitudes and phases.  
This solution has the maximal degree of predictability, as it only depends on one single real parameter, 
the subtraction constant $a$ that merely represents the overall normalization of the amplitude (the phase of which
is of course unobservable); the complete Dalitz plot distribution is then a prediction.

We will discuss various sources of the theoretical errors in this representation,
most of which are in one way or the other associated with the high-energy behavior, in some detail in Sect.~\ref{subsec:error}.
It is obvious, though, that while our high-energy constraints on the amplitudes, including e.g.\ 
the assumed smooth continuation of the scattering phase, present a plausible and internally consistent procedure
to interpolate between the very well-constrained low-energy part and the asymptotic behavior as suggested by the 
Froissart bound, we certainly neglect various details in the description of an intermediate-energy range, in particular
inelastic contributions.  The hope (which, eventually, has to be checked phenomenologically) is that inside the 
dispersive integrals, this intermediate-energy range does not influence the low-energy description of the 
decay amplitudes too much.  
However, similar to what we explained for the Omn\`es function, 
in order to suppress the influence of inelastic contributions even further,
we can alternatively subtract the dispersive solution once more
than strictly necessary, at the expense of introducing another subtraction constant:
\begin{equation}\label{eq:oversub}
\F(s)=\Omega(s)\biggr\{a + b's + \frac{s^2}{\pi}\int_{4\mpc^2}^{\infty}\frac{ds'}{s'^2}\frac{\sin\delta(s')\hat\F(s')}{|\Omega(s')|(s'-s)}\biggr\}~.
\end{equation}
This is equivalent to Eq.~\eqref{eq:inteeqn} if $b'$ obeys the sum rule
\begin{equation}\label{eq:bsumrule}
b' = \frac{1}{\pi}\int_{4\mpc^2}^{\infty}\frac{ds'}{s'^2}\frac{\sin\delta(s')\hat\F(s')}{|\Omega(s')|} ~.
\end{equation}
Because of the special analytic structure of $\hat\F(s)$ discussed in Appendix~\ref{app:struct}, which is due to three-particle cuts
in the decay amplitude, the subtraction constant $b'$ is complex.  If one allows it to take values different from the sum rule
in order to give the dispersive representation more freedom in a fit to experimental data, it therefore represents two new \emph{real}
parameters, modulus and phase of $b'$.  Again, as in the discussion for the Omn\`es function, such a  $b'$ different from its
sum-rule value is in principle inconsistent and leads to a high-energy behavior violating the Froissart bound.  We will, however,
again find that these violations do not manifest themselves in practice in the energy range considered here.

The linearity of Eq.~\eqref{eq:oversub} in the subtraction constants $a$ and $b'$ also massively simplifies the numerical solution
strategy in this case.\footnote{We are grateful to Gilberto Colangelo, Stefan Lanz, and Heiri Leutwyler for pointing this out to us.}
The full solution can be constructed as the linear combination
\begin{equation}
\F(s) = a\big[\F'_a(s) + b \F_b(s) \big] ~,
\label{eq:Fab}
\end{equation}
where $b=b'/a$, in terms of the basis solutions found from 
\begin{align}
\F'_a(s)=\Omega(s)\biggr\{1 + \frac{s^2}{\pi}\int_{4\mpc^2}^{\infty}\frac{ds'}{s'^2}\frac{\sin\delta(s')\hat\F'_a(s')}{|\Omega(s')|(s'-s)}\biggr\}~,\nnnl
\F_b(s)=\Omega(s)\biggr\{s + \frac{s^2}{\pi}\int_{4\mpc^2}^{\infty}\frac{ds'}{s'^2}\frac{\sin\delta(s')\hat\F_b(s')}{|\Omega(s')|(s'-s)}\biggr\}~.\label{eq:Fab-basis}
\end{align}
$\F'_a(s)$ and $\F_b(s)$ can therefore be calculated by the iterative procedure explained above, \emph{before} adjusting the subtraction constants
in a fit to experimental data.

\section{Numerical results}\label{sec:numres}

In this section we show the numerical results from solving Eq.~\eqref{eq:inteeq}. 
We start off by discussing the numerical input in Sect.~\ref{subsec:numinput} before showing the convergence behavior for 
both $\omega\to3\pi$ and $\phi\to3\pi$ in the iteration procedure in Sect.~\ref{subsec:conv}. 
In Sect.~\ref{subsec:Dalitz} Dalitz plot distributions are presented along with a study of crossed-channel 
rescattering effects, followed by 
a discussion on how these effects hold up against the expected errors of our analysis in Sect.~\ref{subsec:error}. 
The comparison to experiment follows in Sect.~\ref{sec:expcomp}.

\subsection{Numerical input}\label{subsec:numinput}

The integral equation Eq.~\eqref{eq:inteeq} is fully determined except for the $\pi\pi$ input and the subtraction constant.  
This input is subject to uncertainties, the discussion of which we defer to Sect.~\ref{subsec:error}.
Instead, we only give a central set of parameters used in the following.
It was already pointed out that we use the partial decay width 
to fix the subtraction constant (that serves as the overall normalization of the amplitude), namely~\cite{PDG}
\begin{equation}
  \Gamma_{\omega\to3\pi}=7.56\,{\rm MeV}~,\qquad\Gamma_{\phi\to3\pi}=0.65\,{\rm MeV}~.
\end{equation}
It should be noted that we do not consider errors on the partial decay widths, 
since the uncertainties thus generated are by far superseded by the error sources discussed in Sect.~\ref{subsec:error}. 
Furthermore, the masses involved are given by $M_\omega = 782.65$\,MeV, $M_\phi=1019.46$\,MeV, and
$\mpc=139.57$\,MeV.
We use the $\pi\pi$ P-wave phase shift
based on an ongoing Roy-equation analysis~\cite{CapriniWIP}
(first aspects of which have recently been published~\cite{CCLRegge}).
There are other parameterizations of the P-wave phase shift of
comparable accuracy available, see for example Ref.~\cite{Pelaez}. 
We will discuss the influence of the difference between the two in our error discussion.

\begin{table}
\centering
\renewcommand{\arraystretch}{1.3}
\begin{tabular}{c c}
\toprule
$\Lambda_\Omega$ & $ 1.3$\,GeV				\\
$\Lambda$ & $ 2.0$\,GeV					\\
\#\,subtractions in $\Omega(s)$ & one			\\
phase-shift param. & Ref.~\cite{CapriniWIP}\\
inelasticities & none\\
\bottomrule
\end{tabular}
\renewcommand{\arraystretch}{1.0}
\caption{Input to the analysis, see text for explanations.}\label{tab:input}
\end{table}

There are other error sources that relate to the upper limit of the dispersion integrals 
in Eqs.~\eqref{eq:Omnesint} and~\eqref{eq:inteeq}. Our procedure in the Omn\`es 
integral is as follows: the phase shift derived from the Roy-equation analysis is strictly known up to 
the validity limit of the Roy equations of  $\sqrt{s}=1.15$\,GeV. 
We use the somewhat extended phenomenological parameterization of Ref.~\cite{CapriniWIP} up to $\Lambda_\Omega=1.3$\,GeV, 
set the phase to a constant beyond that point, and calculate the Omn\`es integral
analytically. 
The upper limit of the dispersion integral in Eq.~\eqref{eq:inteeq} has less physical significance,
it is rather an indicator how well one sums up the remainder of the integral. 
In our analysis the integral is cut off at $\Lambda=2$\,GeV.
We emphasize that this does not mean that we know the physics of $\pi\pi$ interactions up to that point, but it certainly 
is a better approximation to the integrand than setting it to zero.

We have also considered methods to include elastic resonances between 1.3 and 2.0\,GeV ($\rho'(1450)$ and $\rho''(1700)$),  
as well as inelasticities e.g.\ from $4\pi$ intermediate states. Furthermore we have estimated the possible 
contribution of a $\rho_3$-dominated $\pi\pi$ F~wave. 
The corrections stemming from these contributions are tiny and also deferred to the error discussion and the Appendices.
Our input parameters are summarized in Table~\ref{tab:input}.

\subsection{Convergence behavior of the amplitude}\label{subsec:conv}

\begin{figure}
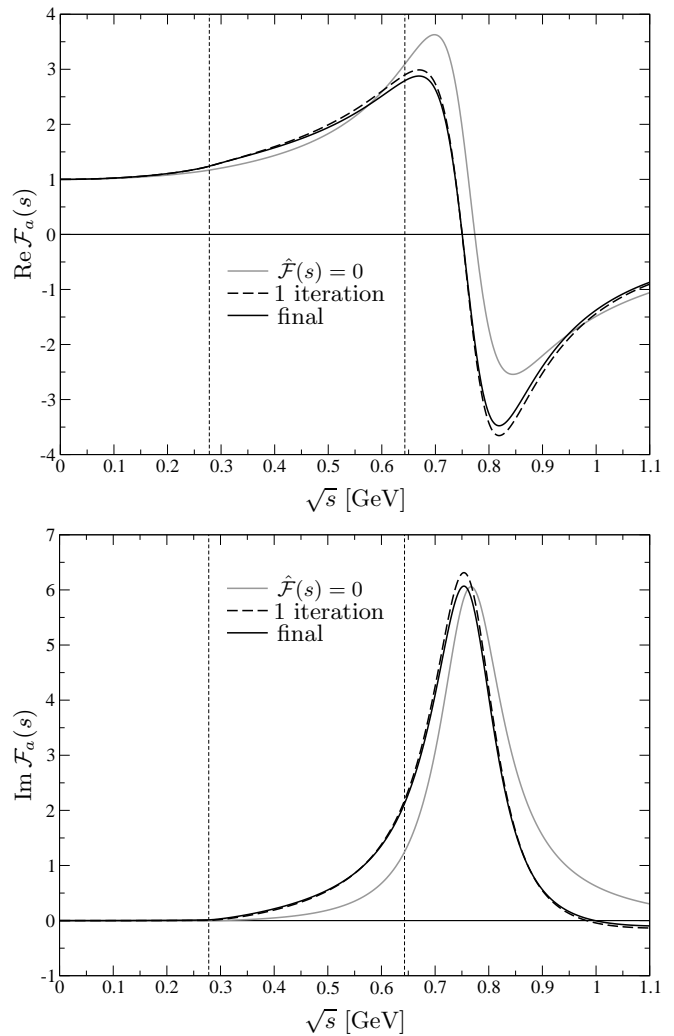

 \centering 
\psfrag{sqrt(s) [GeV]}[c][b][1]{$\sqrt{s}~{\rm[GeV]}$}
\psfrag{Fhat=0}[c][][0.9]{~~$\hat\F(s)=0$}
\psfrag{1 iteration}[c][][1.0]{1 iteration}
\psfrag{final}[c][][1.0]{final}
\psfrag{Re F(s) [a. u.]}[c][][1.0]{$\Re\F_a(s)$}
\psfrag{Im F(s) [a. u.]}[c][][1.0]{$\Im\F_a(s)$}
\includegraphics[width= 0.98\linewidth]{ReFOmegaIt.eps} \\[3mm]
\includegraphics[width= 0.98\linewidth]{ImFOmegaIt.eps}\\[2mm]
 \caption{Successive iteration steps of real (upper panel) and imaginary (lower panel) part of the amplitude $\F_a(s)$ for
$\omega\to 3\pi$. The vertical dashed lines denote the physical region of the decay.}
\label{fig:convomega}
\end{figure}

\begin{figure}
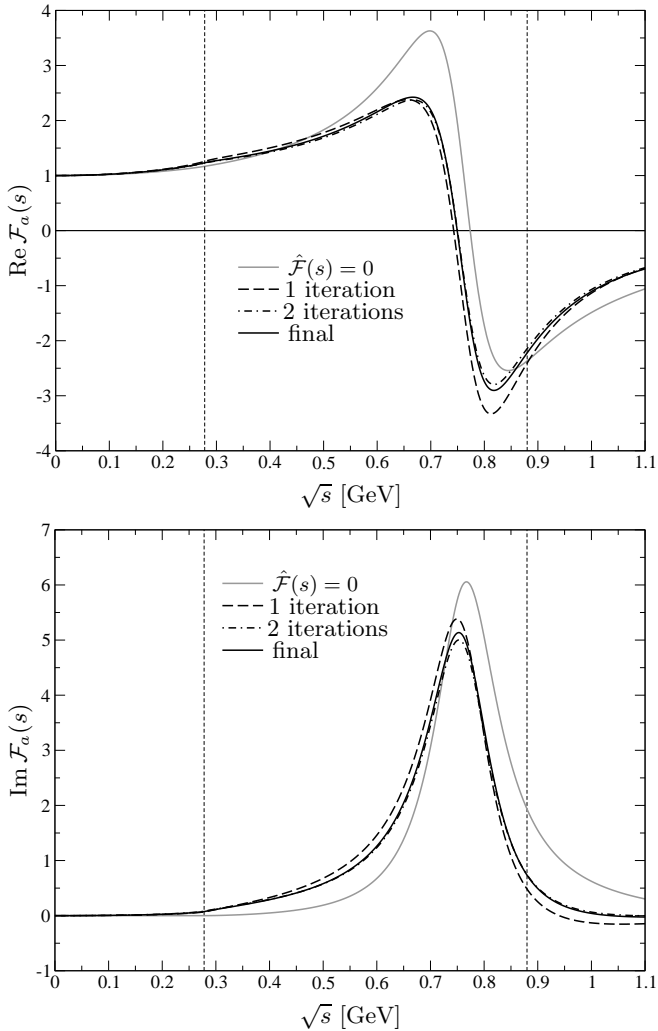

 \centering 
\psfrag{sqrt(s) [GeV]}[c][b][1]{$\sqrt{s}~{\rm[GeV]}$}
\psfrag{Fhat=0}[c][][0.9]{~~$\hat\F(s)=0$}
\psfrag{1 iteration}[c][][1.0]{1 iteration}
\psfrag{2 iterations}[c][][1.0]{2 iterations}
\psfrag{final}[c][][1.0]{final}
\psfrag{Re F(s) [a. u.]}[c][][1.0]{$\Re\F_a(s)$}
\psfrag{Im F(s) [a. u.]}[c][][1.0]{$\Im\F_a(s)$}
\includegraphics[width= 0.98\linewidth]{ReFPhiIt.eps}\\[3mm]
\includegraphics[width= 0.98\linewidth]{ImFPhiIt.eps}\\[2mm]
 \caption{Successive iteration steps of real (upper panel) and imaginary (lower panel) part of the amplitude $\F_a(s)$ for
$\phi\to 3\pi$. The vertical dashed lines denote the physical region of the decay.}
\label{fig:convphi}
\end{figure}

To illustrate the convergence behavior of the iteration procedure, the amplitudes $\F(s)$ for $\omega\to3\pi$ and $\phi\to3\pi$ are plotted after each iteration step in Figs.~\ref{fig:convomega} and~\ref{fig:convphi}.
Note that we show the \emph{normalized} amplitudes, i.e.\ $a=1$. Convergence of the $\omega\to3\pi$ amplitude is 
reached fast, with $\F(s)$ all but indistinguishable from the final result after two iterations. 
The same holds true for $\phi\to3\pi$, although one more iteration step is required.
However, in both cases the difference between the final result and the starting point is significant: 
since our starting point is the Omn\`es function that resums rescattering between two pions only,
this hints at sizeable crossed-channel effects in the decay region.

We observe that the $\rho$ peak in $\omega\to3\pi$ is slightly enhanced and shifted toward lower energies. 
Even though the peak is not part of the physical decay region and this shift therefore not observable directly,
the somewhat steeper rise should leave its imprint on the Dalitz plot.
The $\phi\to3\pi$ amplitude exhibits a similar shift of the $\rho$ peak toward smaller energies. 
In contrast to the $\omega$ decay, the peak here is attenuated, which should certainly have an impact on the Dalitz 
plot distribution, since it is part of the physical region. 
We reiterate that the modifications we observe due to three-particle rescattering lead to non-negligible effects on the amplitude.
In the following section we shall discuss how our observations translate to the actual Dalitz plot distributions.

\subsection{Dalitz plot distributions and crossed-channel\\\hspace*{5mm} rescattering}\label{subsec:Dalitz}

\begin{figure}
  \includegraphics*[width= 0.95\linewidth]{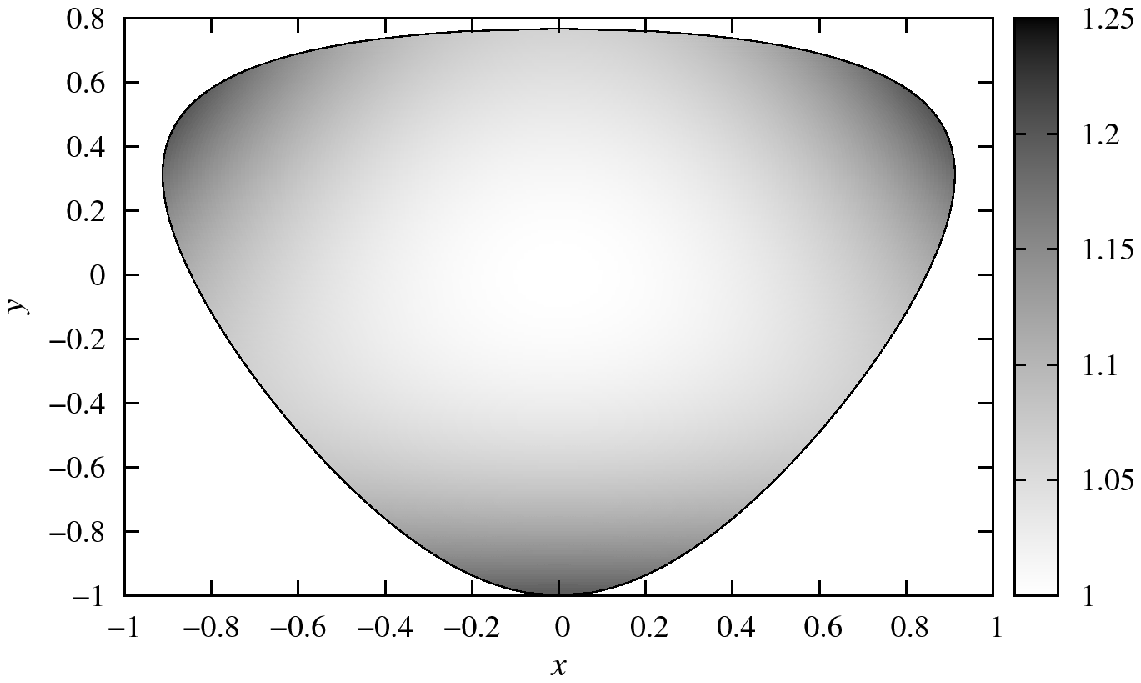}\\[3mm]
  \includegraphics*[width= 0.95\linewidth]{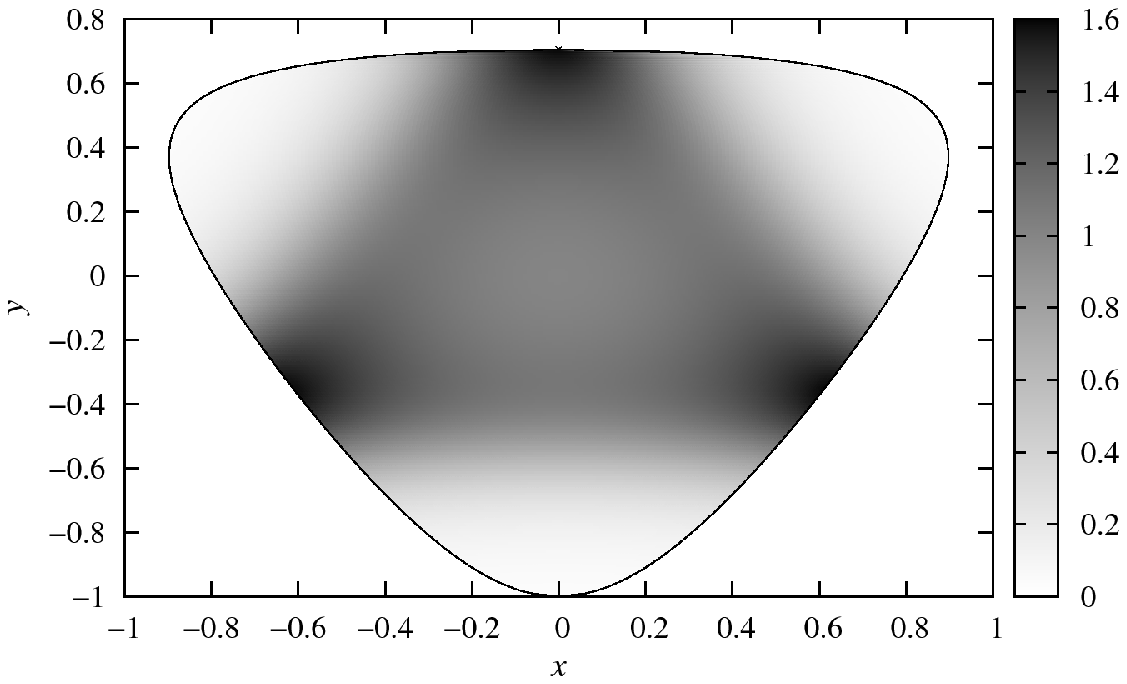}\\[1mm]
  \caption{Dalitz plots for $\omega\to3\pi$ (upper panel) and $\phi\to3\pi$ (lower panel), normalized by the P-wave phase space.}
\label{fig:Dalitzomegaphi}
\end{figure}

For a three-body decay, the Dalitz plot distribution is of particular interest. The Dalitz plot is a 
two-dimensional scatter plot in terms of two kinematic variables; we use the dimensionless variables 
\begin{equation}\label{Dalitzvariable}
x=\frac{t-u}{\sqrt{3}R_V} ~, \quad
y=\frac{s_0-s}{R_V} ~,
\end{equation}
where $R_V=\frac{2}{3}M_V(M_V-3\mpc)$.

In Fig.~\ref{fig:Dalitzomegaphi} we show the Dalitz plot distribution divided by the P-wave phase-space factor given in Eq.~\eqref{eq:phasesp} and normalized to 1 in the center of the Dalitz plot. 
Figure~\ref{fig:Dalitzomegaphi} is thus a pure prediction: 
it is free from any input aside from the $\pi\pi$ P-wave phase shift, 
which is well-established up to at least $1.15$\,GeV, thus covering the entire
physical range for both processes. 

The $\omega\to3\pi$ Dalitz plot exhibits a relatively smooth distribution, 
which rises from the center to its outer borders with a maximum increase of roughly 20\% with respect to
the center. The available phase space is not sufficient to contain the $\rho$ resonance.
This behavior unambiguously fixes the sign of the leading slope parameter in a possible Dalitz plot parameter representation 
to be \emph{positive}, see Sect.~\ref{subsec:Dalrep} for a detailed discussion and numerical results.

The $\phi\to3\pi$ Dalitz plot in contrast shows significantly more structure, since the physical region encompasses 
the $\rho$ resonance: the resonance bands of the $\rho^0$ and $\rho^{\pm}$ are clearly
visible in Fig.~\ref{fig:Dalitzomegaphi}. 
From its center, the Dalitz plot distribution rises towards these bands with a maximum enhancement of roughly 60\% at the peak, 
and then steeply falls off, showing almost complete depletion towards the outer corners.

In order to illustrate effects of crossed-channel rescattering on the Dalitz plot distribution, 
we study the phase-space corrected Dalitz plot after the iteration procedure ($|\F_{\rm full}|^2$), divided by the same quantity 
before the iterations, corresponding to $\hat\F=0$, i.e.\ the sum of pure Omn\`es solutions ($|\F_{\hat\F=0}|^2$). 
We devise two approaches of fixing the subtraction constants. 
First we assume that the subtraction constant is given by some independent method, 
and we are interested in what bearings the crossed-channel effects have on both the overall shape of the Dalitz plot and the partial 
decay width. The quantity $|\F_{\rm full}|^2/|\F_{\hat\F=0}|^2$ is then independent of the specific choice of the subtraction constant. 
For the sake of the argument
we fix the subtraction constant from the experimental decay width for $\hat\F=0$ and then run the iteration procedure. 
Our second approach is to readjust the subtraction constant in such a way as
to reproduce the experimental decay width in both cases, with and without crossed-channel rescattering effects included. 
The focus then lies exclusively on changes to the profile of the Dalitz plot distribution.

\begin{figure}
\centering
\includegraphics*[width = \linewidth]{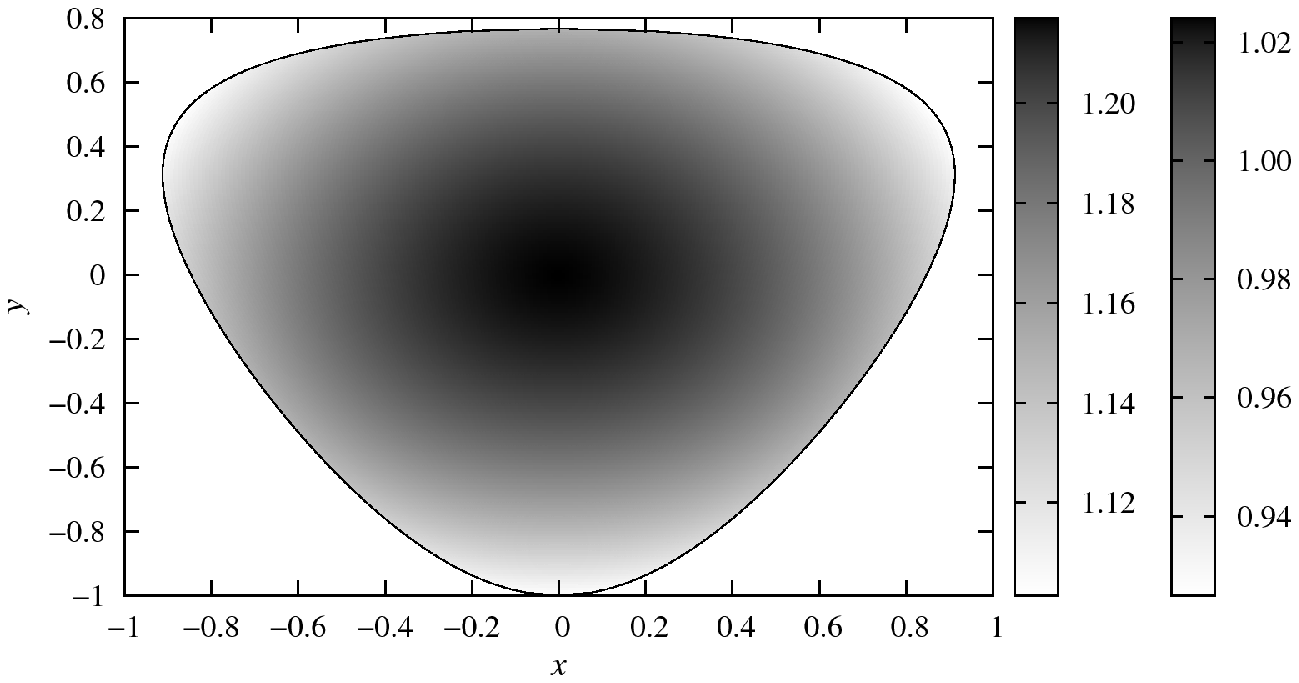} \\[3mm]
\includegraphics*[width = \linewidth]{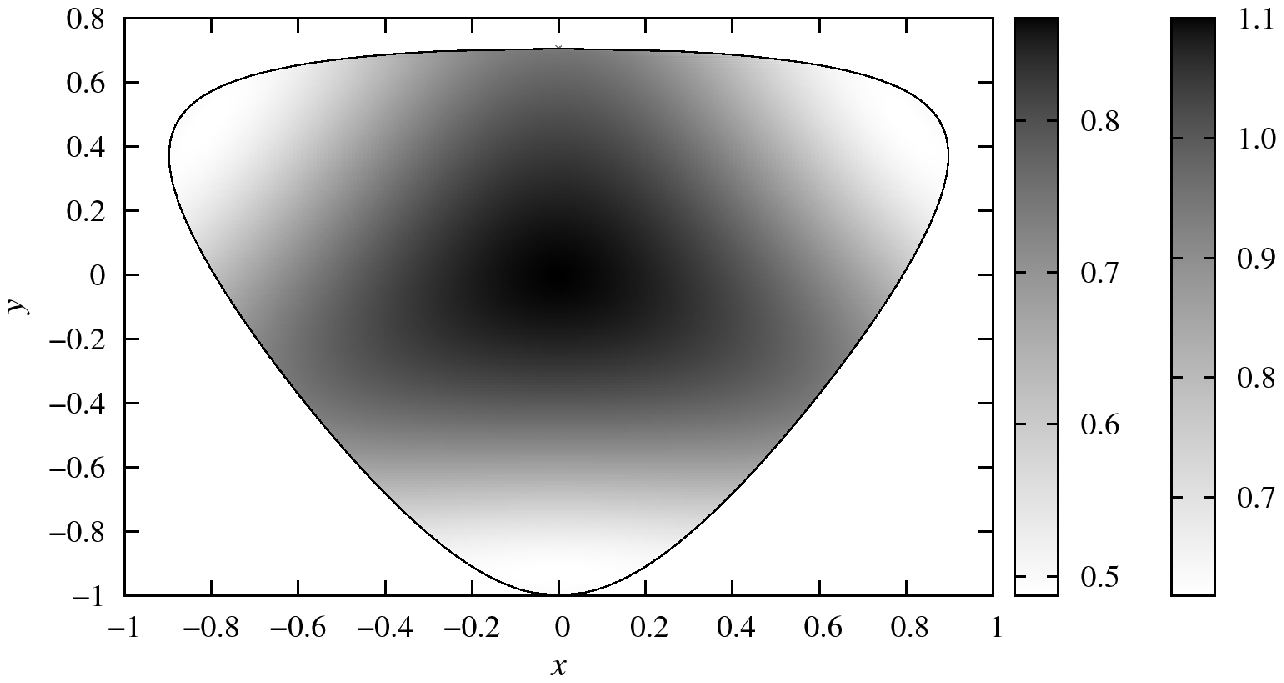} 
\caption{$|\F_{\rm full}|^2/|\F_{\hat\F=0}|^2$ for $\omega\to3\pi$ (upper panel) and $\phi\to3\pi$ (lower panel). 
The two different scales next to the plots correspond to
two different methods of fixing the subtraction constants: 
$a$ is fixed to reproduce the decay width before the iteration only (left scales), or it is
fixed to reproduce the decay width before and after the iteration (right scales).}
\label{fig:crossomegaphi}
\end{figure}
The results of both approaches are shown in Fig.~\ref{fig:crossomegaphi}. 
By keeping the subtraction constant fixed we observe that 
crossed-channel rescattering enhances the $\omega\to3\pi$ partial width by roughly 20\%. 
This qualitative behavior translates to the Dalitz plot, where the enhancement of the distribution ranges between
10\% on the borders and 24\% in the center. 
The $\phi\to3\pi$ partial width in contrast sees a decrease by likewise roughly 20\%. 
The decrease across the Dalitz plot is stronger, with 60\% on the border,
than in the center where it amounts to roughly 10\%.
Fixing the subtraction constant before \emph{and} after the iteration procedure, 
the effects due to crossed-channel rescattering are to a large degree absorbed in the partial width in both decays.
Indeed, the remaining effect on the $\omega\to3\pi$ Dalitz plot amounts to an 8\% decrease that is alleviated 
towards the central region, where one observes a slight increase of roughly 3\%. The same qualitative
behavior is observed in $\phi\to3\pi$, only quantitatively stronger: 
the 50\% suppression on the border is counterbalanced by a 20\% enhancement in the center. 
The $\rho$ bands are left unscathed by the iteration procedure.

\begin{sloppypar}
Overall, we find that crossed-channel rescattering leaves a significant imprint on the Dalitz plot distribution. 
However, before checking how our approach to those three-particle effects holds up against 
experimental scrutiny, we study whether the size of the effects even withstands the uncertainties of our input parameters 
in the following section.
\end{sloppypar}

\subsection{Error discussion}\label{subsec:error}

\begin{table}
\centering
\renewcommand{\arraystretch}{1.3}
\begin{tabular}{c c c}
\toprule
				& $\omega\to3\pi$ [\%] & $\phi\to3\pi$ [\%]\\
\midrule
\#\,subtractions 		& $5.2$ 		& $4.5$ 	\\
$\delta(s)$ parameterization	& $2.4$			& $2.6$		\\
$\Lambda_\Omega$ 		& $0.6$			& $0.8$		\\
$\delta(s)\to\pi$		& $0.4$			& $0.6$		\\
inelasticities 	 		& $0.2$			& $0.4$		\\
\bottomrule
\end{tabular}
\renewcommand{\arraystretch}{1.0}
\caption{Relative variation of $\Im\F(s)$ in the $\rho$ peak as percentage of our standard input parameters 
(Table~\ref{tab:input}) for the respective error source for $\omega\to3\pi$ (left) and $\phi\to3\pi$ (right). See text for further
details.}\label{tab:error}
\end{table}

As we stated in the previous sections our results are---aside from the subtraction constant---fully constrained by the 
$\pi\pi$ P-wave phase shift. The parameterizations of the phase shift we use are very accurate
in the low-energy regime; in this section we study the influence of uncertainties that are mainly due to the high-energy behavior and different methods to assess it.
These error sources are listed in Table~\ref{tab:error} together with the maximum variation they produce in $\Im\F(s)$ 
on the $\rho$ peak relative to our central parameter set (see Table~\ref{tab:input}). In the following, 
we discuss these error sources in more detail.

As mentioned in Sect.~\ref{subsec:FF}, using a twice-subtracted Omn\`es function is a means to suppress the high-energy behavior
of the phase-shift input. 
If we naively plug our central parameter set into Eq.~\eqref{eq:sumrule} we obtain 
\begin{equation}
\langle r_\text{sum}^2 \rangle^V_\pi\simeq 0.415\, {\text{fm}}^2~, \label{eq:Nsumrule}
\end{equation}
which lies somewhat below a next-to-next-to-leading order chiral perturbation theory (ChPT) analysis~\cite{BijTal},
\begin{equation}\label{eq:scalrad}
\langle r_{\rm ChPT}^2 \rangle^V_\pi= 0.452\pm 0.013\, {\text{fm}}^2~,
\end{equation}
and the current particle-data-group average~\cite{PDG}.
Dispersive analyses of $e^+e^-\to\pi^+\pi^-$ data~\cite{Bern:piFF1,Bern:piFF2} point towards a value much closer to Eq.~\eqref{eq:Nsumrule}, 
with central values of the order of $\langle r^2 \rangle^V_\pi\simeq 0.43\,{\text{fm}}^2$;
however, we consider the variation between a once-subtracted and twice-subtracted Omn\`es function, 
using the phenomenological radius in the latter,  a conservative estimate of the uncertainty. 
We find that the variation of the $\rho$ peak amounts to roughly 5\% and is thus the largest source
of uncertainty.

It is essential to try to assess the uncertainties stemming from our main input, the parameterization of the $\pi\pi$ P-wave 
phase shift. 
Since these phase-shift solutions in themselves are very accurate, especially in the low-energy regime, 
we address these errors by varying between two different parameterizations,
referred to as ``Bern''~\cite{CapriniWIP} and ``Madrid''~\cite{Pelaez} parameterizations in the following.
These lead to a difference in $\Im\F(s)$ in the $\rho$ peak by roughly 2\%, 
which amounts to the second largest error in our analysis.
We have also checked how higher resonances ($\rho'(1450)$ and $\rho''(1700)$) 
can modify the phase shift (beyond the range of the Roy analyses) and hence our results. 
This aspect is described in greater detail in Appendix~\ref{app:highres}; the impact on the decay amplitudes
in the physical region however turns out to be negligible.

\begin{sloppypar}
Related to the previous point is the exact value of $\sqrt{s}$ beyond which the phase is set to a constant, $\Lambda_\Omega$. 
To check the dependence on this error, we vary $\Lambda_\Omega=1.3\ldots1.5$\,GeV. 
Note that the phase shift beyond $1.3$\,GeV is not well constrained at all, 
yet it still produces a very small uncertainty below 1\%
and will be neglected in the following. 
An equally negligible modification is found when leading the phase continuously to $\pi$ beyond $\Lambda_\Omega$ instead of setting
it to a constant, which is in principle in better agreement with the $1/s$ 
behavior of the $\pi\pi$ vector form factor for large $s$.
\end{sloppypar}

\begin{figure}
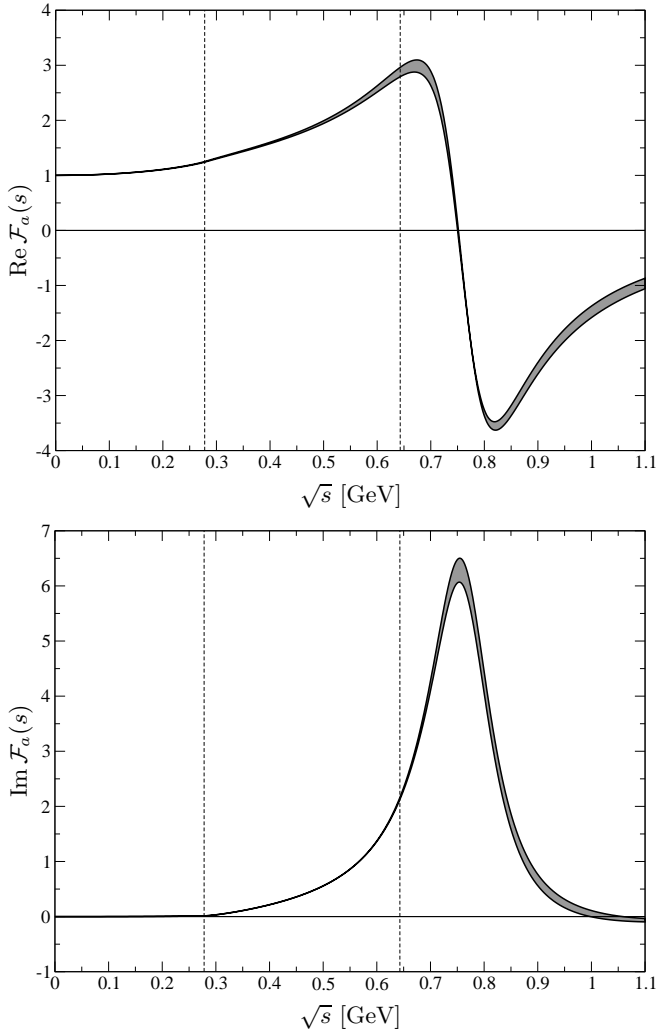

 \centering 
\psfrag{sqrt(s) [GeV]}[c][b][1]{$\sqrt{s}~{\rm[GeV]}$}
\psfrag{Re F(s) [a. u.]}[c][][1.0]{$\Re\F_a(s)$}
\psfrag{Im F(s) [a. u.]}[c][][1.0]{$\Im\F_a(s)$}
\includegraphics[width= 0.98\linewidth]{ReFOmegaerr.eps} \\[3mm]
\includegraphics[width= 0.98\linewidth]{ImFOmegaerr.eps} \\[2mm]
 \caption{Real (upper panel) and imaginary (lower panel) part of the amplitude $\F_a(s)$ for
$\omega\to 3\pi$. The vertical dashed lines denote the physical region of the decay. 
The shaded line represents the uncertainty generated by upper and lower boundary of our solution (see text).}
\label{fig:omegaerr}
\end{figure}

Inelasticities in the $\pi\pi$ P~wave have been incorporated in a simplified fashion, following
the method proposed in Ref.~\cite{Walker},
by bestowing the exponential of the phase shift with an inelasticity parameter $\eta(s)\doteq \eta_1^1(s)$,
\begin{equation}\label{eq:inelpara}
 t_1^1(s)=\frac{\eta(s)e^{2i\delta(s)}-1}{2i}~.
\end{equation}
This leads to modified dispersion integrals (see Ref.~\cite{Walker} and Appendix~\ref{app:inpara}),
\begin{align}\label{eq:inteeqinhom}
\F(s)&=a \,\xi(s)\Xi(s)\Omega(s) \nnnl
&\times\!\biggl\{1 + \frac{s}{2\pi i}\int_{4\mpc^2}^{\infty}\frac{ds'}{s'}\frac{\big[e^{i\delta(s')}-\eta(s')e^{-i\delta(s')}\big]\hat\F(s')}{\sqrt{\eta(s')}\Xi(s')|\Omega(s')|(s'-s)}\biggr\},
\end{align}
where
\begin{align}
\xi(s)= \begin{cases}
 \eta^{-1/2}(s) &\text{above the cut,}\\
 \eta^{1/2}(s) &\text{below the cut,}\\
 1 &\text{elsewhere,}
\end{cases}
\end{align}
and
\begin{equation}
 \Xi(s)=\exp\biggl\{\frac{is}{2\pi}\dashint{10pt}\limits_{16\mpc^2}^\infty ds'\frac{\log\eta(s')}{s'(s'-s)}\biggr\}~,
\end{equation}
where $\dashint{0pt}$ denotes the principal value integral, and we assume that inelasticities set in at the four-pion threshold. 
The inelasticity starts showing major deviations from unity only above roughly $1$\,GeV (see also Sect.~\ref{sec:inel}). 
Consequently the effects of using such a simplified model for the inelasticity show little impact (below half a percent) 
on our final result. 

We add one final remark on the integral cutoff in the dispersion integral of the full amplitude, Eq.~\eqref{eq:inteeq}. 
This cutoff has been fixed to 2\,GeV and \emph{not} varied. The reason for this is
that it is merely a ``numerical'' cutoff: the kinematic range of validity is fixed by the $\pi\pi$ phase-shift parameterization.
The dispersion integral has no physical content beyond that point; we just 
need to ensure that the integral cutoff is large enough so as not to produce numerical artifacts.
We found that this is the case at 2\,GeV. 
As a side remark, we have checked that lowering the integral cutoff as far as to 1.3\,GeV still affects the amplitudes
in the physical region by less than the main sources of uncertainty discussed above.

\begin{figure}
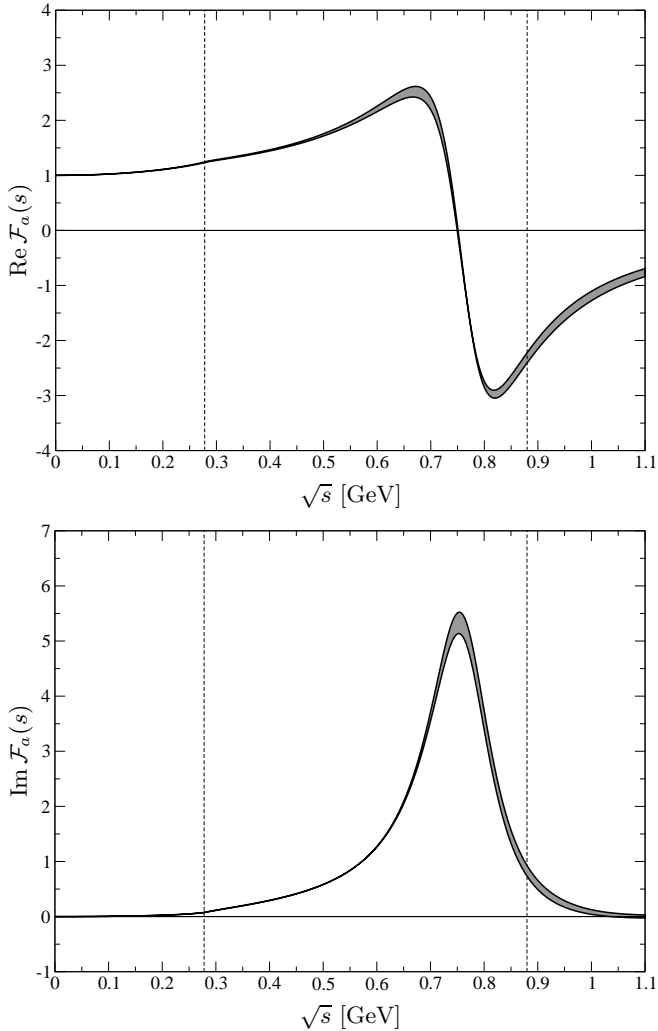

 \centering 
\psfrag{sqrt(s) [GeV]}[c][b][1]{$\sqrt{s}~{\rm[GeV]}$}
\psfrag{Re F(s) [a. u.]}[c][][1.0]{$\Re\F_a(s)$}
\psfrag{Im F(s) [a. u.]}[c][][1.0]{$\Im\F_a(s)$}
\includegraphics[width= 0.98\linewidth]{ReFPhierr.eps}\\[3mm]
\includegraphics[width= 0.98\linewidth]{ImFPhierr.eps}\\[2mm]
 \caption{Real (upper panel) and imaginary (lower panel) part of the amplitude $\F_a(s)$ for
$\phi\to 3\pi$. The vertical dashed lines denote the physical region of the decay. 
The shaded line represents the uncertainty generated by upper and lower boundary of our solution (see text).}
\label{fig:phierr}
\end{figure}

After identifying the charge radius in the twice-sub\-tracted Omn\`es function and the phase-shift parameterization 
as the main \emph{single} error sources, we will now perform a more detailed 
study of their combined error. 
It turns out that we can identify certain parameter sets as upper and lower boundaries of the amplitude for that purpose. 
The combination of the Bern parameterization along with a once-subtracted Omn\`es function will serve as the lower boundary, 
whereas the upper boundary is given by the Madrid phase together with a twice-subtracted Omn\`es function and the 
determination of the pion charge radius from Ref.~\cite{BijTal}. 
The uncertainty bands generated between these boundaries are shown for $\omega\to3\pi$ and $\phi\to3\pi$ in 
Figs.~\ref{fig:omegaerr} and \ref{fig:phierr}. The maximum uncertainty of 
the amplitude in the physical region amounts to roughly 8\% in both cases.

To rule out that interference of the crossed channels amplifies the uncertainties, 
we also analyze how the errors develop across the Dalitz plot. For that purpose, we study the following quantity across
the Dalitz plot:
\begin{equation}
 \frac{|\F_u|^2-|\F_l|^2}{|\F_u|^2+|\F_l|^2}~,
\end{equation}
where $\F_{u/l}$ denotes the normalized amplitude on the upper/lower boundary. 
We find that for $\omega\to3\pi$ the error in the Dalitz plot is even slightly decreased with respect to the amplitude, 
the maximal uncertainties amount to 1.4\% towards the corners of the Dalitz plot.
For $\phi\to3\pi$ we observe the largest uncertainties in the outer edges of the Dalitz plot, 
where the Dalitz plot strength itself is strongly suppressed. So while the relative error there rises up to 8\%, the 
absolute error is very small.

Comparing to our results in Sect.~\ref{subsec:Dalitz} we clearly see that crossed-channel contributions are sizable enough 
to outweigh the uncertainties. In the following section we will analyze whether
crossed-channel effects are actually observable in experiment.

\section{Comparison to experiment}\label{sec:expcomp}
\subsection{Fit to the \boldmath{$\phi\to 3\pi$} Dalitz plot: single subtraction}

We now wish to compare our theoretical predictions to the experimental results for the $\phi\to 3\pi$ Dalitz plot
measurements by the KLOE~\cite{KLOE} and CMD-2~\cite{CMD-2} collaborations.
The former has significantly larger statistics (almost $2\times 10^6$ events in the Dalitz plot) than the latter
(close to $8\times 10^4$ events); furthermore, the energy resolution in the KLOE measurement was significantly better
than the bin size (in the range $1$--$2\,$MeV compared to a bin size of $8.75\,$MeV), such that smearing effects were found to be negligible, and we could
fit our amplitudes to efficiency-corrected data directly (with purely statistical errors based
on data and Monte Carlo statistics), while for the comparison with the CMD-2 data, they had
to be convoluted with efficiency matrices by the collaboration before.  
While consistency with both data sets is clearly desirable, we will present the comparison to the KLOE data in some more
detail in the following.

Our first goal is to perform a fit to the Dalitz plot distribution with our most predictive theoretical representation, 
Eq.~\eqref{eq:inteeq}, employing a single subtraction, such that
the normalization is the only free parameter of the fit. The shape of the Dalitz plot is thus
purely a prediction, and we can compare the $\chi^2$ of the fit with and without crossed-channel rescattering. 
There are two caveats to this procedure, which we need to discuss beforehand.

First, our calculations are performed in the isospin limit of equal charged and neutral pion masses; 
we use the \emph{charged} pion mass, not least for consistency reasons due to the fact 
that the $\pi\pi$ phase shifts are only available in the isospin limit, with the charged pion mass used 
as the reference quantity.
The effect of this approximation on the amplitude is expected 
to be small (compare e.g.\ Ref.~\cite{etaDal} for a detailed study in the context of $\eta\to3\pi$), with
the main difference due to different charged and neutral pion masses showing up in the available phase space: 
the true physical Dalitz plot is slightly larger than in our calculation. 
To account for this dominant isospin-breaking correction, 
we therefore multiply the amplitude (squared) with the physical phase-space factor:
\begin{equation}
 |\M_{\phi\to3\pi}(s,t,u)|^2=\frac{s}{16}\kappa_{0}^2(s)\sin^2\theta_0|\F(s,t,u)|^2~, \label{eq:phasespN}
\end{equation}
where in contrast to Eq.~\eqref{eq:phasesp} we have  $\cos\theta_0=(t-u)/\kappa_0(s)$ and
\begin{equation}
 \kappa_{0}(s)=\sqrt{1-\frac{4M_{\pi^\pm}^2}{s}}\lambda^{1/2}(M_\phi^2,\mpn^2,s)~.
\end{equation}
Numerically, we employ $\mpn=134.98$\,MeV.
\begin{figure}
\centering
\includegraphics*[width = \linewidth]{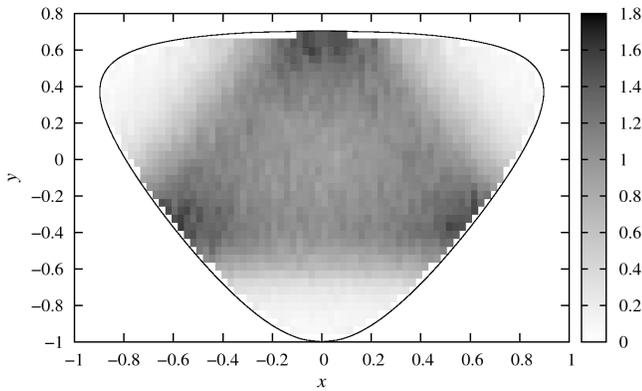}
\caption{Selected data from the KLOE measurement~\cite{KLOE}. 
Shown is the efficiency-corrected number of counts in the respective bin, 
divided by the phase-space factor in Eq.~\eqref{eq:phasespN} and normalized to 1 in the Dalitz plot center.}
\label{fig:KLOEdata}
\end{figure}
Furthermore, in order to avoid distortions due to threshold effects in the decay amplitude, 
we omit data bins that cross the boundary of the Dalitz plot, see Fig.~\ref{fig:KLOEdata},
thereby ensuring that our amplitude is never evaluated below the isospin-symmetric threshold, 
that is, for $(M_{\pi^\pm}+M_{\pi^0})^2 \leq t,\,u \leq 4M_{\pi^\pm}^2$.

\begin{table*}
\centering
\renewcommand{\arraystretch}{1.3}
\begin{tabular}{c c c c c}
\toprule
				& $\F_l$ 	& $\F_u$ 		& $\F_l$ 	& $\F_u$ \\
                                &\multicolumn{2}{c}{(full)}             & \multicolumn{2}{c}{($\hat\F=0$)} \\
\midrule
$\chi^2/{\rm ndof}$ 		& $1.50$ 	& $1.17$ 		& $1.71$		& $2.06$	\\
$\tilde a\times{\rm GeV}^{3}$    & $228.6\pm0.2$ & $216.6\pm0.2$		& $200.2\pm0.2$		& $187.6\pm0.2$	\\
$a_\omega\times 10^3$ 	& $7.9\pm0.5$   & $7.4\pm0.5$  		& $10.9\pm1.0$ 		& $12.6\pm1.2$  \\
$\phi_\omega$ 		& $-0.30\pm0.11$& $-0.10\pm0.11$ 	& $0.87\pm0.06$		& $0.95\pm0.06$ \\
\bottomrule
\end{tabular}
\renewcommand{\arraystretch}{1.0}
\caption{Fit results to the KLOE data for $\phi\to 3\pi$, using the once-subtracted dispersive representation.
Shown are the outcomes of the fits for upper and lower boundaries of the theoretical uncertainty band, with and
without crossed-channel rescattering included. The normalization of $\tilde a$ is arbitrary, such that
only the relative changes in $\tilde a$ between different fits are significant, not the absolute values.
The uncertainties quoted refer to the errors of the fit.}\label{tab:fit}
\end{table*}

\begin{sloppypar}
The second caveat, discussed in Ref.~\cite{KLOE}, concerns the fact that the $\phi$ is produced in $e^+e^-$ collisions
at DA$\Phi$NE,  $e^+e^-\to\phi\to\pi^+\pi^-\pi^0$, which allows for the background process 
$e^+e^-\to\omega\pi^0\to\pi^+\pi^-\pi^0$, the $\omega\pi^0$ invariant mass equalling the mass of the $\phi$, 
with the $\omega$ subsequently decaying into $\pi^+\pi^-$.
As the decay $\omega\to\pi^+\pi^-$ violates isospin, with the branching fraction suppressed to the percent level~\cite{PDG},
the overall modification of the Dalitz plot is small.\footnote{Note that this is an isospin-violating
effect specific for the $3\pi$ production in $e^+e^-$ collisions.  If we interpret the decay
$\omega\to\pi^+\pi^-$ in terms of a $\rho$--$\omega$ mixing angle $\theta_{\rho\omega}$ 
(see e.g.\ Ref.~\cite{Urech}), this effect is \emph{linear} in $\theta_{\rho\omega}$ 
due to the fact that the photon has both isoscalar and isovector components.  In contrast, isospin breaking
due to $\rho$--$\omega$ mixing in $\pi\pi$ scattering is necessarily suppressed to second order
in $\theta_{\rho\omega}$ and hence irrelevant.} 
However, it is entirely concentrated in the narrow band
$s = M_\omega^2$, and leaves a visible effect there.
There are two possible strategies to deal with this issue: one could simply omit the corresponding
horizontal band in a fit of the Dalitz plot; or, alternatively, add a resonance term of the form (see Ref.~\cite{KLOE})
\begin{equation}
 \A_{\omega\pi}(s)=a \times a_\omega e^{i\phi_\omega}\frac{M_\omega^2}{M_\omega^2- i \sqrt{s} \Gamma_\omega-s} 
\end{equation}
to Eq.~\eqref{eq:SVAadd} after the iteration. (Note that we have factored out the normalization constant $a$ for reasons
of comparison with the fit results in Ref.~\cite{KLOE}.) 
$M_\omega$ and $\Gamma_\omega$ are fixed to the particle-data-group values~\cite{PDG}, 
and $a_\omega$ and $\phi_\omega$ are dimensionless fitting parameters. We follow the latter 
strategy, in particular since the resonance term also has a small impact on bins adjacent to the horizontal band at $83.7$\,MeV.
\end{sloppypar}

Our standard $\chi^2$ fit is performed with 1834 data points and three free parameters. 
We perform separate fits for the upper and lower boundaries of the theoretical uncertainty band as discussed in 
Sect.~\ref{subsec:error}, both with and without crossed-channel rescattering effects included.
The results are listed in Table~\ref{tab:fit}.
The data set fitted to is given in arbitrary normalization, such that the constant $\tilde a$ does not 
correspond to the real subtraction constant $a$ and is only shown for comparison of the changes between different fits.
We notice that the fit quality considerably improves once crossed-channel rescattering is taken into account. 
This indicates that crossed-channel effects lead to modifications that are not only non-negligible, but even 
\emph{observable} in the structure of the Dalitz plot. 
It is also observed, however, that the precision of the data is such that the 
variation of the $\chi^2/{\rm ndof}$ within the theoretical 
uncertainty is quite non-negligible:
the upper boundary yields a considerably better fit than the lower boundary 
for the full crossed-channel analysis, while the opposite holds for the pure Omn\`es solutions.
The $a_\omega$ coupling is found to be in the 1\% range, the expected order of magnitude; 
its numerical value determined together with the full
dispersive solution is in slightly better agreement with what is found in Ref.~\cite{KLOE}.
The best $\chi^2/{\rm ndof}$ of 1.17 of our fits is not quite as good as it is in Ref.~\cite{KLOE}, 
and it needs to be pointed out that due to the high number of degrees of freedom, 
the $p$-value characterizing the goodness of the fit is still rather low even in the best case, 
$p(\chi^2=1.17) = 3\!\times\!10^{-7}$.
However, this fit quality is achieved with less degrees of freedom:
apart from the $\omega\pi^0$ background term (which has nothing to do with the genuine $\phi\to3\pi$ Dalitz plot),
this is a one-parameter fit,
the shape of the Dalitz plot in our case is a pure prediction.
In Ref.~\cite{KLOE} there are two additional degrees of freedom by fitting the (complex) ``background'' term. 
From the construction of our dispersive amplitude, it is obvious that in this form, such an independent background
term is inconsistent with the requirements of analyticity and unitarity.
Furthermore, in Appendix~\ref{app:highres} we show that a simplified approach to including higher resonances 
does not allow for an improvement of the fit as opposed to claims made for the nature
of the background term being due to $\rho'$ effects. 
It seems that crossed-channel effects saturate the background term in the KLOE data to a large degree. 

\begin{figure*}
\centering
\includegraphics[width = 0.83\linewidth]{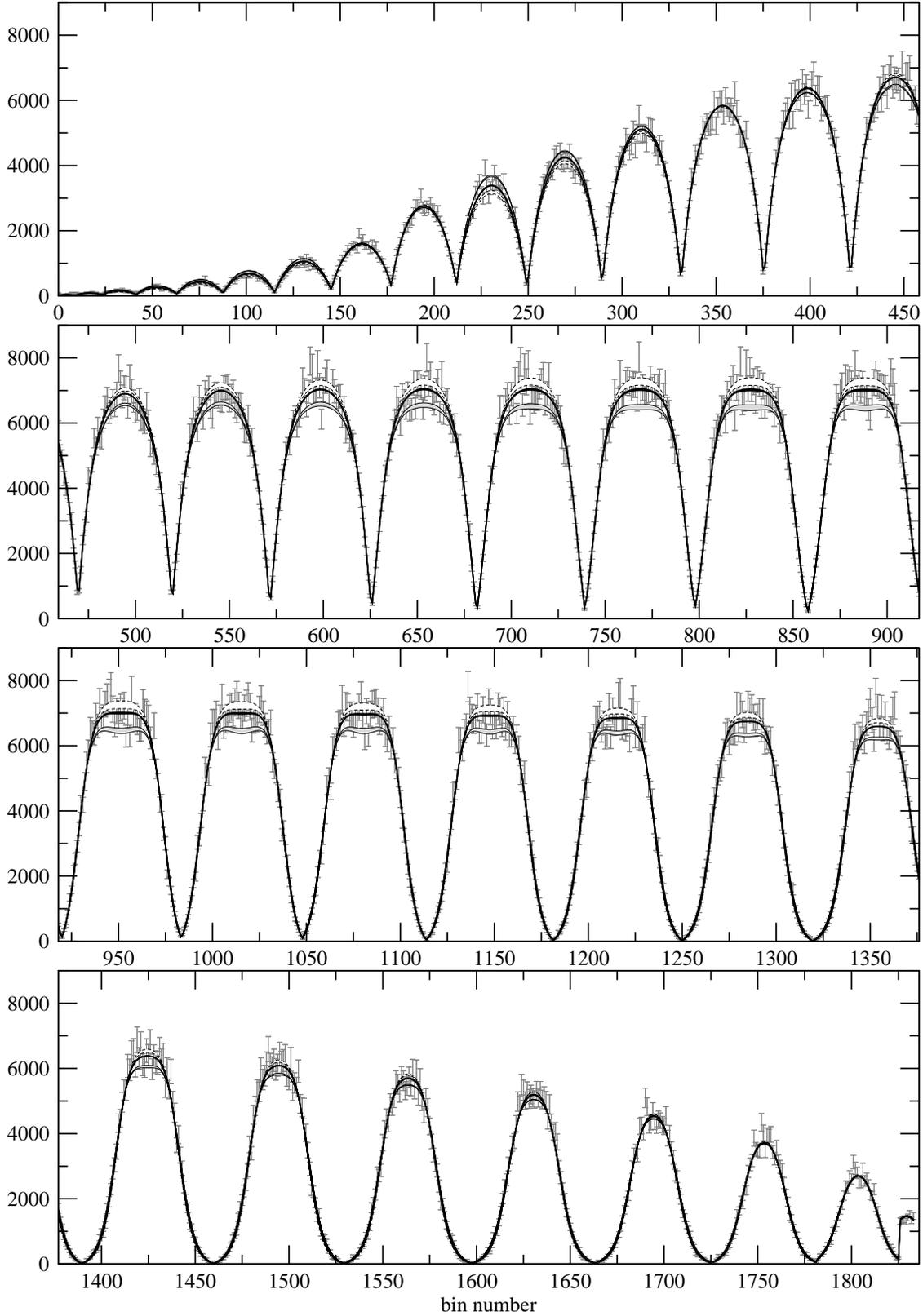}
\caption{Efficiency-corrected number of events per bin plotted against bin number for slices of constant $y$.
Every peak corresponds to one slice/one specific value of $y$, progressing from negative to positive $x$,
and slices are ordered from negative to positive $y$.
Our fit results are plotted for the sum of Omn\`es functions $\hat\F=0$ (gray band), 
for the full solution of the once-subtracted ansatz (white band with dashed boundaries),
and for the full solution in the twice-subtracted variant (black band),
together with the KLOE data~\cite{KLOE}.}
\label{fig:fit}
\end{figure*}
Figure~\ref{fig:fit} shows slices of constant $y$ through the Dalitz plot, 
where the number of events per bin divided by the bin efficiency is plotted against the bin number 
(note that this is not the same as Fig.~5 in Ref.~\cite{KLOE}, where slices of constant $x$ are shown). 
Our fit results are displayed as error bands. The full solution (white band with dashed boundaries) 
gives a better description of the data than the sum of three Omn\`es functions (gray band), 
particularly in the central region of the Dalitz plot. 

These conclusions on the significance of rescattering effects are not immediately substantiated by the 
comparison to the CMD-2 data~\cite{CMD-2}.
The fits performed by the collaboration seem to lead to the almost opposite result:
a fit based on the sum of Omn\`es functions leads to a very good $\chi^2/{\rm ndof}$ 
of about 1.0 for both variants of the Omn\`es function discussed; 
while the full amplitudes yield bad fit qualities of $\chi^2/{\rm ndof} = 1.5 \ldots 1.8$. 
We note, however, that due to the significantly smaller number of degrees of freedom (${\rm ndof}=197$),
the $p$-value for the CMD-2 fit of the full $\F_u$ solution, $p(\chi^2=1.50)=7\!\times\!10^{-6}$, is even better
than the best KLOE fit in Table~\ref{tab:fit}.
Still, this is a surprising result in different respects: primarily, 
as the phenomenological fits of Breit--Wigner $\rho$ resonance terms plus a constant ``background'' amplitude yield perfectly
compatible results in both experiments; secondly, as Ref.~\cite{CMD-2} cites the significance for a non-vanishing background
term at $3.3\sigma$, it is somewhat unexpected that simply replacing the Breit--Wigner $\rho$ by an Omn\`es function
is already sufficient to yield a good description of the data.

\subsection{Fit to the \boldmath{$\phi\to 3\pi$} Dalitz plot: two subtractions}

In order to understand the situation of the two $\phi\to 3\pi$ data sets better, we attempt a description
with the (more flexible) twice-subtracted representation Eqs.~\eqref{eq:Fab}, \eqref{eq:Fab-basis}.
Due to the equivalence with the once-subtracted form in case the additional parameter $b$ fulfils the sum rule
Eq.~\eqref{eq:bsumrule}, the fits can only improve: the hope is to find an acceptable fit also to the CMD-2 data 
that is compatible with the fundamental principles underlying the dispersive representation.
We refrain from employing the two different variants of the Omn\`es function defined earlier; we expect the 
second subtraction in the dispersion integral to render the second subtraction inside the Omn\`es function redundant,
and therefore only use the standard form~\eqref{eq:Omnesint}.
We perform the fit of this representation with the two different $\pi\pi$ P-wave phase 
parameterizations~\cite{CapriniWIP,Pelaez} to check the consistency of the outcome.  

\begin{table}
\centering
\renewcommand{\arraystretch}{1.3}
\begin{tabular}{c c c}
\toprule
				& Bern~\cite{CapriniWIP}	& Madrid~\cite{Pelaez} \\
\midrule
$\chi^2/{\rm ndof}$ 		& $1.02$ 	& $1.03$ 		\\ 
$\tilde a\times{\rm GeV}^3$     & $207.6\pm1.4$ & $207.1\pm1.5$		\\ 
(sum rule)                      & $228.6$       & $225.4$               \\ 
$|b|\times{\rm GeV}^{-2}$        & $0.97\pm0.03$ & $0.94\pm0.03$         \\ 
(sum rule)                      & $0.72$        & $0.75$                \\ 
$\arg b$                        & $0.52\pm0.03$ & $0.42\pm0.03$         \\ 
(sum rule)                      & $0.73$        & $0.70$                \\ 
$a_\omega\times 10^3$ 	& $7.3\pm0.6$   & $7.5\pm0.6$  		\\ 
$\phi_\omega$ 		& $ 0.40\pm0.10$& $ 0.33\pm0.10$ 	\\ 
\bottomrule
\end{tabular}
\renewcommand{\arraystretch}{1.0}
\caption{Fit results to the KLOE data for $\phi\to 3\pi$, using the twice-subtracted dispersive representation.
Shown are the outcomes of the fits using the phase parameterization of the Bern~\cite{CapriniWIP} and
Madrid~\cite{Pelaez} groups.  The ``sum rule'' entries for modulus and phase of $b$ refer 
to the evaluation of Eq.~\eqref{eq:bsumrule} with the once-subtracted dispersive representation;
the ``sum rule'' entry for $a$ just serves as a reminder that also the normalization of the amplitude
changes significantly between the two fits. For $\tilde a$, the same remark holds as in Table~\ref{tab:fit}.
The uncertainties quoted refer to the errors of the fit.}\label{tab:fit2}
\end{table}
The results of the fits to the KLOE data are shown in Table~\ref{tab:fit2}.  The resulting values for $a$ and $b$
are compared to the ones obtained from the fit of the once-subtracted representation, see the previous section,
via the sum rule~\eqref{eq:bsumrule}.  For both phases, excellent fits of $\chi^2/{\rm ndof}$ close to 1.0 are obtained
($p(\chi^2) = 0.22\ldots 0.24$).
Closer comparison shows that $\Im b$ in fact stays very close to its sum-rule value (in particular for
the Bern phase parameterization), while $\Re b$ is shifted (enlarged) more significantly, at the expense of a somewhat
reduced normalization $\tilde a$.
\begin{figure}
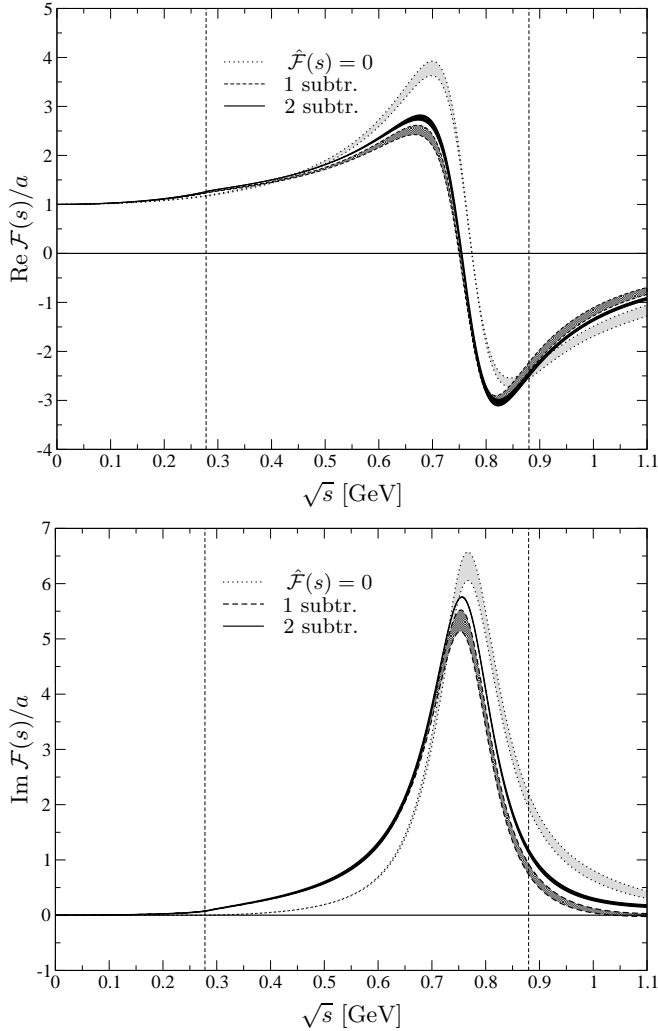

\centering
\psfrag{sqrt(s) [GeV]}[c][b][1]{$\sqrt{s}~{\rm[GeV]}$}
\psfrag{ReF [a. u.]}[c][][1.0]{$\Re\F(s)/a$}
\psfrag{ImF [a. u.]}[c][][1.0]{$\Im\F(s)/a$}
\psfrag{ReOmnes}[c][][0.9]{~~$\hat\F(s)=0$}
\psfrag{ImOmnes}[c][][0.9]{~~$\hat\F(s)=0$}
\psfrag{ReFull1}[c][][0.9]{~~1~subtr.}
\psfrag{ImFull1}[c][][0.9]{~~1~subtr.}
\psfrag{ReFull2}[c][][0.9]{~~2~subtr.}
\psfrag{ImFull2}[c][][0.9]{~~2~subtr.}
\includegraphics[width= 0.98\linewidth]{ReFabfull.eps}\\[3mm]
\includegraphics[width= 0.98\linewidth]{ImFabfull.eps}\\[2mm]
\caption{Real (upper panel) and imaginary (lower panel) part of the amplitude $\F(s)$ (normalized with $a=1$) for
$\phi\to 3\pi$, comparing the fit of the twice-subtracted dispersive representation to the KLOE data
to the once-subtracted $\F(s)$ as well as the Omn\`es solution.  All curves are shown with the uncertainty
bands attached as discussed in the text.}
\label{fig:Fabfull}
\end{figure}
We compare the resulting function $\F_a(s)+b\F_b(s)$ to the once-subtracted $\F(s)$ as well as the Omn\`es function
in Fig.~\ref{fig:Fabfull}.  We see that, indeed, the difference to the once-subtracted $\F(s)$ is small compared
to the latter's error band, let alone the difference to the Omn\`es function.
We also display the fit result in comparison to the KLOE data in 
Fig.~\ref{fig:fit} as the black band.  Note that as we now optimize the additional parameter through the fit routine
and only vary between the two different phase parameterizations, the band is significantly narrower than the other two.
While the deviation from the once-subtracted fit results seems to be really minor, 
the overall improvement in the $\chi^2/{\rm ndof}$ is significant.

Fits to the CMD-2 data have again been performed by the collaboration.  This time, they result in 
a $\chi^2/{\rm ndof}$ of 0.96 and 0.94 (with associated $p$-values of $0.64\ldots 0.71$), 
using the Bern and Madrid phases, 
finding 
\begin{align}
|b| &= \big\{ 0.97^{+ 0.16}_{- 0.13} \,,\, 0.95^{+0.15}_{- 0.12} \big\} ~, \nnnl
\arg b &= \big\{0.00 \pm 0.16 \,,\, -0.18 \pm 0.18 \big\} ~, 
\end{align}
respectively.  
While the modulus of $b$ therefore agrees perfectly with the fit to the KLOE data, 
the phase seems to prefer a \emph{real} $b$, in contradistinction to the sum-rule prediction.
Only the KLOE data therefore yield a high significance for a non-vanishing phase of the parameter $b$.
We have not investigated systematic uncertainties in the determination of $b$ from data.

\subsection{\boldmath{$\omega\to3\pi$} Dalitz plot parameterization}\label{subsec:Dalrep}

Since there is currently no precise data available on the $\omega\to3\pi$ Dalitz plot, we will now discuss the issue 
whether it is feasible to use a parameterization of the Dalitz plot in terms of a polynomial
in a precision determination, and if so, with how many terms necessary.
This is common practice for decays with final-state particles at low energies, 
such as $K\to3\pi$, $\eta\to3\pi$, or $\eta'\to\eta\pi\pi$.
Since the Dalitz plot for $\omega\to\pi^+\pi^-\pi^0$ is relatively smooth, 
we expect that a similar description should be possible here, even though the phase space available is somewhat larger. 
Moreover, as opposed to e.g.\ the decays
$K^\pm\to\pi^0\pi^0\pi^\pm$, $\eta\to3\pi^0$, and $\eta'\to\eta \pi^0\pi^0$, one does not face the issue of cusp 
effects~\cite{CGKR,GKR,BFGKR,etaprime} 
in the decay region that give rise to possibly large non-analytic structures.

The $\omega\to3\pi$ decay amplitude is fully symmetric under exchange of
$s$, $t$, and $u$, see Eq.~\eqref{eq:SVAadd}, 
such that the Dalitz plot description is formally similar to the one in $\eta\to3\pi^0$. 
For a parameterization in terms of a polynomial,
it is therefore convenient to rewrite the Dalitz plot variables in Eq.~\eqref{Dalitzvariable} 
in polar coordinates, namely
\begin{equation}\label{eq:xybar}
 y \doteq\sqrt{z}\sin\phi~, \quad
 x \doteq\sqrt{z}\cos\phi~.
\end{equation}
We can therefore parameterize the amplitude squared in terms of $z$ and $\phi$ according to
\begin{align}
 |\F_{\rm pol}(z,\phi)|^2=|\N|^2\Big\{1&+2\alpha z+2\beta z^{3/2}\sin3\phi+2\gamma z^2\nnnl
&+2\delta z^{5/2}\sin{3\phi}+\Order\big(z^3\big)\Big\}~,
\end{align}
where $\N$ is the normalization and $\alpha$, $\beta$, $\gamma$, $\delta$ are the Dalitz plot parameters, 
in strict analogy to $\eta\to 3\pi^0$~\cite{etaDal}.
Note in particular that we attempt to parameterize $|\F|^2$ in polynomial form, not $|\M|^2$:
the trivial kinematic factor due to the P-wave characteristic of the vector-meson decay should always be retained exactly.
In the following, we study whether such a parameterization makes sense when trying to quantitatively describe 
the $\omega\to3\pi$ Dalitz plot, and give predictions for the parameters.

We perform a fit of the polynomial Dalitz plot representation $|\F_{\rm pol}(z,\phi)|^2$ to our theoretically determined amplitude 
$|\F_{\rm th}(z,\phi)|^2$, minimizing the function
\begin{align}
\chi^2 &= \frac{1}{\N_\mathcal{D}}\int_{\mathcal{D}}dzd\phi  \biggl[
\bigg( 1 - \frac{3z(3s_0-R_\omega\sqrt{z}\sin3\phi)}{(M_\omega+3M_\pi)^2} \bigg) \nnnl
& \qquad\qquad\qquad \times \frac{|\F_{\rm pol}(z,\phi)|^2-|\F_{\rm th}(z,\phi)|^2}{|\N|^2}\biggr]^2~, \nnnl
\N_\mathcal{D} &= \int_\mathcal{D}dzd\phi ~,\label{eq:chi2Dal}
\end{align}
where $\mathcal{D}$ denotes the area of the Dalitz plot.
The normalization of Eq.~\eqref{eq:chi2Dal} is chosen such that $\sqrt{\chi^2}$ denotes the average
deviation of the polynomial parameterization in $|\M(s,t,u)|^2$, relative to the Dalitz plot center.
The prefactor (in round brackets) corresponds to the kinematic factor relating $|\M(s,t,u)|^2$ to
$|\F(s,t,u)|^2$ in Eq.~\eqref{eq:phasesp}, rewritten in terms of $z$ and $\phi$, and normalized to 1 
in the center of the Dalitz plot.
We decide to include the kinematic phase-space factor in the minimization in order to give less weight
to the outer parts of the Dalitz plot, which are expected to also contribute statistically less in an experimental determination.

\begin{table*}
\centering
\renewcommand{\arraystretch}{1.3}
\begin{tabular}{c c c c c c c c}
\toprule
			& $|\N|\times{\rm GeV}^3$& $\alpha\times10^3$	& $\beta\times10^3$	& $\gamma\times10^3$ & $\delta\times10^3$ & $\sqrt{\chi^2}\times10^3$		&max dev.$\times10^3$\\
\midrule																
full 			& $1451\ldots1447$	& $84\ldots96\,(102)$ 	& --- 			& ---	   		& ---	      		&  $0.9\ldots 1.1$		&$40$\\
			& $1453\ldots1449$	& $74\ldots84\,(90)$	& $24\ldots28\,(30)$	& ---	   		& ---			&  $0.052\ldots 0.078$		&$3$\\
			& $1453\ldots1449$	& $73\ldots81\,(86)$	& $24\ldots28\,(29)$	& $3\ldots6\,(8)$	& ---			&  $0.038\ldots 0.047$		&$3$\\
			& $1453\ldots1449$	& $74\ldots83\,(88)$	& $21\ldots24\,(25)$	& $0\ldots2\,(3)$	& $7\ldots8\,(9)$	&  $0.012 \ldots 0.011$		&$2$\\
\midrule																		
$\hat\F=0$ 		& $1433\ldots1429$	& $137\ldots148$ 	& --- 			& ---	   		& ---	      		&  $1.1\ldots 1.3$		&$50$\\
			& $1435\ldots1431$	& $125\ldots135$	& $29\ldots33$		& ---	   		& ---			&  $0.25\ldots 0.29$		&$20$\\
			& $1436\ldots1433$	& $113\ldots120$	& $26\ldots29$		& $24\ldots 27$		& ---			&  $0.036\ldots 0.045$		&$4$\\
			& $1436\ldots1433$	& $114\ldots122$	& $23\ldots25$		& $20\ldots 23$		& $7\ldots8$		&  $0.002\ldots 0.003$		&$0.3$\\
\bottomrule
\end{tabular}
\renewcommand{\arraystretch}{1.0}
\caption{Fit results for the Dalitz plot parameters with (above) and without (below) crossed-channel effects included. 
We show the Dalitz plot parameters along with the values for $\sqrt{\chi^2}$ as defined in the text.
``max dev.'' is the maximum deviation at any point across the Dalitz plot between the full solution and the polynomial fit.
The numbers in brackets for the full solution refer to the extension of the parameter ranges considering 
the twice-subtracted dispersion relation; for details, see main text.}\label{tab:fitresults}
\end{table*}

We start with the singly subtracted dispersive representation, where the Dalitz plot shape is a full-fledged prediction.
Our results are summarized in Table~\ref{tab:fitresults}, where again we vary the amplitude between upper and lower error boundary
of the once-subtracted solution, and also compare to the pure Omn\`es solution without rescattering effects.
We observe that for the full amplitude, a fit with two parameters (the normalization is not counted as a fit quantity) 
already gives a very good description of the theoretical data: compared to the 
one-parameter fit, the $\sqrt{\chi^2}$ 
is improved by roughly a factor of 15, the maximum relative deviation across the Dalitz plot, 
i.e.\ the quantity $|\F_{\rm pol}(z,\phi)|^2/|\F_{\rm th}(z,\phi)|^2-1$ at any point,
is improved by approximately a factor of $10$ from 4\% to 0.3\%. 
When increasing the number of Dalitz plot parameters beyond 2, the changes in the parameters themselves as well as 
the quality of the fit are rather small, and quite probably beyond the reach even of a precision experiment. 
Note that the maximum deviation between two and three parameters is approximately the same, 
however the value of $\sqrt{\chi^2}$ is significantly reduced. The reason for this is that the contributions with stronger
deviation are pushed toward the outer boundary of the Dalitz plot, owing to the phase-space factor in Eq.~\eqref{eq:chi2Dal}.

The Omn\`es solution shows a slightly different behavior: a good fit quality is reached only with three parameters, 
and the third parameter $\gamma$ is found to be significantly larger.
Deviations across the Dalitz plot with only two terms are still substantial at 2\% and probably within reach
of an experimental determination. Also the parameters $\alpha$ and $\beta$ still see changes in the 10\% range.
The comparison of our predictions of the Dalitz plot parameters between the full and the Omn\`es solution 
hints at a significant influence of crossed-channel rescattering. Note that, beyond the precise values of the Dalitz plot
parameters, the sign of the leading parameter $\alpha$ is unambiguously fixed in $\omega\to3\pi$ as discussed before
(see Sect.~\ref{subsec:Dalitz}).

Given our comparison to the experimental $\phi\to 3\pi$ Dalitz plots, we may still wonder how reliable
these predictions of $\omega\to3\pi$ Dalitz plot parameters are.  It is obvious that as soon as we switch 
to the twice-subtracted dispersive representation, we cannot strictly predict all of these any more: at least
$\alpha$ would have to be an input quantity.  In order to estimate the potential effects, we resort to the 
following procedure: we assume the deviations of the second subtraction constant $b$ from the sum-rule result 
to be moderate; as these should be due to imperfectly understood high-energy behavior of the amplitudes, 
we take the \emph{relative} deviation in the corresponding $\phi\to3\pi$ subtraction constant as an upper limit
on what we deem acceptable for $\omega\to3\pi$.  The sum-rule values for $b$, see Eq.~\eqref{eq:bsumrule}, in 
$\omega\to3\pi$ are
\begin{equation}
b_\text{sum} = \big\{ 0.54 \,e^{0.14i} \,,\, 0.56 \,e^{0.13i} \big\}
\end{equation}
for the Bern and Madrid $\pi\pi$ phase solutions, respectively.  
Note that the phase $\arg b$ is significantly smaller for the $\omega$
decay compared to the $\phi$: the imaginary part in the subtraction constant is a three-particle-cut effect and as such
proportional to the phase space available for the three pions in the corresponding decays.  
Enlarging $b$ by the same factors as required in the fits to the KLOE $\phi\to3\pi$ data, compare Table~\ref{tab:fit2},
we evaluate the twice-subtracted $\omega\to3\pi$ amplitude with 
\begin{equation}
b=\big\{0.83 \,e^{0.09i} \,,\, 0.83 \,e^{0.07i} \big\} 
\end{equation}
instead.  The results for the Dalitz plot parameters tend to lie only slightly above the ranges of variation found 
within the uncertainty band for the once-subtracted representation; they are quoted in brackets for the full solution
in Table~\ref{tab:fitresults}.

We conclude that we need at least two parameters to obtain a polynomial parameterization of the $\omega\to3\pi$
Dalitz plot at the 1\% level accuracy, and thus directional information (owing to the $\phi$-dependence) is required. 
Let us compare this situation to the $\eta\to3\pi^0$ Dalitz plot, which has the same three-fold symmetry and therefore
a Dalitz plot distribution that is almost flat.  The slope parameter is now measured to excellent precision to be
$\alpha(\eta\to3\pi^0) = -0.0315\pm0.0015$~\cite{PDG}; we predict $\alpha(\omega\to3\pi)$ to be about 2.5 times as large,
and of opposite sign.  No higher-order Dalitz parameters have ever been determined for $\eta\to3\pi^0$; a theoretical
calculation predicts 
$\beta (\eta\to3\pi^0) = (-4.2\pm0.7)\times 10^{-3}$, 
$\gamma(\eta\to3\pi^0) = ( 1.3\pm0.4)\times 10^{-3}$~\cite{etaDal},
hence these terms beyond the linear term in $z$ will modify the Dalitz plot density only at the few-permille level
($z\leq 1$).  
In contrast, $\beta(\omega\to3\pi)$ is larger than $\beta (\eta\to3\pi^0)$ 
by almost an order of magnitude, see Table~\ref{tab:fitresults}, and 
hence expected to be significantly more important/more likely to be determined experimentally.

As a final remark on potential uncertainties in the Dalitz plot parameters, 
we note that in an experimental investigation of $\omega\to3\pi$, the invariant mass of the detected three pions is 
going to vary within the natural width of the $\omega$.  
We have checked that taking this variation into account in the calculation of the decay amplitude
the two leading Dalitz plot parameters $\alpha$ and $\beta$ change at the permille level, 
hence way below the level of uncertainty of our prediction, 
and probably also significantly below the accuracy of any experimental determination in the near future.
Note that $M_\omega$ is {\em not} changed in the definition of the Dalitz plot variables, Eq.~\eqref{eq:xybar}.

\section{\boldmath{$\pi\pi$} P-wave inelasticity}\label{sec:inel}

\begin{figure}
\centering
\includegraphics[width=0.55\linewidth]{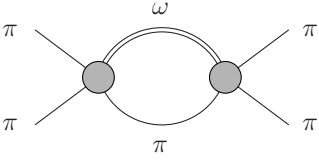}
\caption{Contribution to the $\pi\pi$ P-wave inelasticity due to $\omega\pi$ intermediate states.
Each gray blob corresponds to an $\omega\to3\pi$ amplitude, analytically continued to the scattering region.}
\label{fig:inelasticityCut}
\end{figure}
It is well known that $K\bar K$ intermediate states play the dominant role in the generation of inelastic
effects in the $\pi\pi$ isospin $I=0$ S~partial wave, where the $K\bar K$ threshold almost coincides with the 
position of the $f_0(980)$ resonance, up to at least 1.3\,GeV.  In contrast, the contribution of 
$K\bar K$ to the $\pi\pi$ $I=1$ P-wave inelasticity $\eta_1^1$ is almost negligible, 
and $\eta_1^1$ is believed to be dominated by $4\pi$ contributions.  Phenomenologically, one finds
that the onset of this inelasticity happens roughly at $\sqrt{s} \simeq M_\omega+\mpc\simeq 0.922$\,GeV, such that it 
is a natural question to ask whether $\omega\pi$ intermediate states
as an effective two-body description of four pions give an adequate description of the 
$\pi\pi$ P-wave inelasticity, at least in a certain energy region above threshold. 
This requires knowledge of the inelastic scattering amplitude $\pi\pi\to\omega\pi$ in the P~wave---precisely
the crossed process of $\omega\to3\pi$.
The inelasticity parameter $\eta_1^1$ can be calculated from the cut contribution shown in Fig.~\ref{fig:inelasticityCut},
yielding
\begin{equation}
 \eta_1^1(s)=\sqrt{1-\frac{q_{\pi\pi}^3(s)q_{\omega\pi}^3(s)}{144\pi^2}\big|f_1(s)\big|^2\theta(s-(M_\omega+\mpc)^2)}~,
\end{equation}
where $q^2_{ab}(s)=\lambda(s,M_a^2,M_b^2)/4s$, and $f_1(s)=\F(s)+\hat\F(s)$ 
is the P-wave projection of the $\pi\pi\to\omega\pi$ amplitude.

\begin{figure}
\centering
\psfrag{eta(s)}{$\eta_1^1(s)$}
\psfrag{sqrt(s) [GeV]}{$\sqrt{s}$ [GeV]}
\includegraphics[width = 0.94\linewidth]{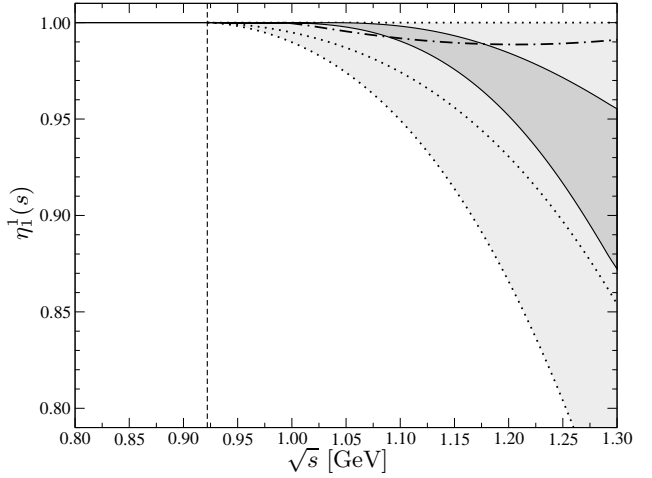}
\caption{Inelasticity in the $\pi\pi$ P~wave as a function of $\sqrt{s}$. 
The dark gray band represents the contribution of $\omega\pi$ intermediate states (within our theoretical uncertainty);
the light gray band shows the error range of the phenomenological determination of Ref.~\cite{CapriniWIP},
with the central value and error limits given by the dotted curves.
The dash-dotted curve shows the inelasticity contribution of the $K\bar K$ intermediate state for comparison.
The vertical dashed line denotes the $\omega\pi$ threshold.}
\label{fig:inelasticity}
\end{figure}
The resulting inelasticity is shown in Fig.~\ref{fig:inelasticity}, up to $\sqrt{s} = 1.3$\,GeV, 
and for the uncertainty band in the $\omega\to3\pi$ amplitude discussed in Sect.~\ref{subsec:error}.
It is seen that this error grows rapidly with energy.
Phenomenological determinations of $\eta_1^1(s)$ are also plagued by rather large uncertainties~\cite{CapriniWIP,Pelaez};
Fig.~\ref{fig:inelasticity} shows that the $\omega\pi$ intermediate state seems to provide most of the inelasticity
given as the central value of the analysis in Ref.~\cite{CapriniWIP}.  There is seemingly a difference in the threshold
behavior, with the $\omega\pi$ contribution rising more slowly near threshold; this is due to the fact that our 
$\pi\pi\to\omega\pi$ partial wave $f_1(s)$ happens to have almost a zero around 1\,GeV.
We hasten to add that $\omega\pi$ is not the only way to cluster four pions into an effective
two-body P-wave state; other possibilities like $\rho\sigma$ would just be expected to set in
at an even higher effective mass.

For comparison, we also plot the inelasticity contribution from $K\bar K$ intermediate states,
derived from the parameterization of the $\pi\pi\to K\bar K$ partial wave $g_1^1(s)$ given in Ref.~\cite{piK} 
(see there for definitions), based on the data from Ref.~\cite{KKdata}.
This yields a $\pi\pi$ P-wave inelasticity contribution according to
\begin{equation}
 \eta_1^1(s)_{K\bar K}=\sqrt{1-\frac{16}{s}q_{\pi\pi}^3(s)q_{KK}^3(s)\big|g_1^1(s)\big|^2\theta(s-4M_K^2)}~,
\end{equation}
which, in Fig.~\ref{fig:inelasticity}, is indeed seen to be very small, and actually remains so 
up to at least $\sqrt{s}\simeq 2$\,GeV.

\section{Summary and conclusions}\label{sec:summary}

In this article we have performed a dispersive analysis of the decays $\omega\to3\pi$ and $\phi\to3\pi$. 
Our framework allows for a  treatment of crossed-channel two-body rescattering effects fully consistent
with analyticity and unitarity.
It contains the three-particle cuts generated by $\mV>3\mpc$, thus going fundamentally beyond 
an isobar-like description.
We have shown that crossed-channel rescattering produces a significant effect on the Dalitz plot 
clearly exceeding uncertainties generated by the phenomenological input for
the P-wave phase shift, discontinuities in higher partial waves, 
and elastic P-wave resonances other than the $\rho(770)$.

Comparing to the very precise measurements of the $\phi\to3\pi$ Dalitz plot by the KLOE collaboration,
we have found indications for these rescattering effects. Indeed, we obtain
a considerably improved $\chi^2$ with the full dispersion calculation 
with one subtraction compared to an isobar-like description by Omn\`es functions. 
Performing a fit of a more flexible, yet less predictive twice-subtracted dispersion calculation,
we can achieve an excellent description with high probability of both 
the KLOE and the CMD-2 data sets, with the more precise KLOE data suggesting a significant 
deviation of the second subtraction constant from its sum-rule value.

For $\omega\to3\pi$ we give predictions for the Dalitz plot parameters for the full dispersion calculation 
as well as neglecting crossed-channel effects. A precise measurement of these parameters that
is to be expected from upcoming experimental analyses by KLOE and WASA should be able to distinguish 
between both approaches. Moreover we have observed that a polynomial fit to the $\omega\to3\pi$ 
Dalitz plot at percent-level precision should require at least two parameters, in contrast
e.g.\ to the decay $\eta\to 3\pi^0$ that displays a similarly symmetric Dalitz plot.
The sign of the leading polynomial term is constrained by the $\rho$ peak to be positive.

The inelasticity of the $\pi\pi$ P~wave is dominated by $4\pi$ intermediate states, 
which can be effectively approximated by a two-body description as $\omega\pi$.
We have found that the inelasticity obtained from the analytic continuation of our
$\omega\to 3\pi$ amplitude  
is in reasonable agreement with the not very well-constrained phenomenological determination.

Finally, we wish to remark that it has become amply clear in the course of this article that the treatment
of both decays, $\omega\to3\pi$ and $\phi\to3\pi$, runs strictly in parallel and shows no formal difference
other than the mass of the decaying vector meson, and the overall normalization of the Dalitz plot, 
which relates to the (partial) decay widths.  Obviously, the very same formalism can also be applied 
to the Dalitz plot description of $e^+e^-\to3\pi$ at arbitrary invariant mass of the electron--positron pair
$\sqrt{s_{e^+e^-}}$, which replaces the vector mass $M_V$.  It only has to be kept in mind that this would 
involve an $s_{e^+e^-}$-dependent normalization---dispersion theory does not allow us to predict the 
energy-dependence of the $e^+e^-\to3\pi$ total cross section.  Given that we have shown that crossed-channel
rescattering effects differ in a non-trivial way for the $\omega$ and $\phi$ decays, it may still be interesting
to investigate the $s_{e^+e^-}$-dependence of these corrections in a systematic way.
Ultimately, for such a task it might be useful to strive for a combination of dispersion theory with microscopic models
that can provide parameterizations for the energy-dependent subtraction constants.

\begin{acknowledgement}
\textbf{Acknowledgements} \ We are indebted to the KLOE and CMD-2 collaborations for invaluable
help and discussions on the data comparison, 
in particular to C.~Bini, S.~Giovannella, A.~Kup\'s\'c, 
S.~I.~Eidelman, D.~A.~Epifanov, and B.~A.~Shwartz;
C.~Bini for providing us with direct access to the data of Ref.~\cite{KLOE},
D.~A.~Epifanov for performing fits of our amplitudes to the data of Ref.~\cite{CMD-2},
and A.~Kup\'s\'c for originally suggesting to generalize our study to $\phi\to 3\pi$.
We would like to thank 
G.~Colangelo for providing us with the preliminary $\pi\pi$ phase-shift results of Ref.~\cite{CapriniWIP},
and P.~Roig for e-mail communication and fit parameters concerning Ref.~\cite{Roig}.
Furthermore, we are grateful to G.~Colangelo, M.~Hoferichter, A.~Kup\'s\'c, S.~Lanz, and S.~Leu\-pold for numerous
useful discussions, and U.-G.~Mei{\ss}ner for comments on the manuscript.
Partial financial support by
the DFG (SFB/TR 16, ``Subnuclear Structure of Matter''),
by the project ``Study of Strongly Interacting Matter'' 
(HadronPhysics2, Grant Agreement No.~227431 and HadronPhysics3, Grant Agreement No.~283286) 
under the 7th Framework Program of the EU,
and by the Bonn--Cologne Graduate School of Physics and Astronomy
is gratefully acknowledged.
\end{acknowledgement}

\appendix

\section{On the size of higher partial waves}\label{app:Fwave}

In this Appendix, we investigate the potential uncertainty in our amplitude solution due 
to the omission of discontinuities in higher partial waves for $\omega\to 3\pi$, the next higher
one being the F~wave.  We wish to emphasize to begin with that higher partial waves do not
vanish for the solution $\F(s,t,u)$: projecting onto the F~wave (in the $s$-channel)
\begin{equation}
f_3(s) = -\frac{7}{16}\int_{-1}^1 dz_s \big(5z_s^4-6z_s^2+1\big)\F(s,t,u) ~,
\end{equation}
where $z_s = \cos\theta_s = (t-u)/\kappa(s)$, 
yields contributions from the $t$- and $u$-channel P-wave amplitudes.  
$f_3(s)$ thus calculated just happens to be \emph{real}.  
To generalize our approach and also include F-wave discontinuities, we may amend the 
decomposition~\eqref{eq:SVAadd} according to
\begin{align}
\F(s,t,u) &= \F(s)+\F(t)+\F(u) \nnnl & + P'_3(z_s) \G(s) + P'_3(z_t) \G(t) + P'_3(z_u) \G(u) ~,
\label{eq:defP+Famp}
\end{align}
where $\G(s)$ again only has a right-hand cut, 
and $z_t=(s-u)/\kappa(t)$, $z_u=(s-t)/\kappa(u)$.
The discontinuities of P and F~wave are expressed by the relations
\begin{align}
\disc f_1(s) &= \disc\F(s) \nnnl
&= 2i\big( \F(s) + \hat\F(s)\big)\theta(s-4M_\pi^2)\sin\delta_1^1(s)e^{-i\delta_1^1(s)} \,, \nnnl
\disc f_3(s) &= \disc\G(s) \nnnl
&= 2i\big( \G(s) + \hat\G(s)\big)\theta(s-4M_\pi^2)\sin\delta_3^1(s)e^{-i\delta_3^1(s)} \,,
\end{align}
where $\delta_3^1(s)$ is the $\pi\pi$ F-wave phase shift, 
and the inhomogeneities are now given by
\begin{align}
\hat\F(s) &= 3 \langle (1-z_s^2) (\F+P'_3\G)\rangle(s) ~, \nnnl
\hat\G(s) &= -\frac{7}{4}\langle \big(5z_s^4-6z_s^2+1\big)(\F+P'_3\G) \rangle(s)  ~. \label{eq:Fangav}
\end{align}
Note that, in Eq.~\eqref{eq:Fangav}, the notation of angular averaging $\langle\ldots\rangle$
is generalized compared to Eq.~\eqref{eq:inhomogen} in the sense that the argument of $P'_3$
is taken to be $z_t$, reexpressed in terms of $s$ and $z_s$.
We will now proceed to estimate the potential size of the F-wave discontinuity contribution
$\G(s)$ by calculating the effects of the $\rho_3(1690)$ resonance.

\subsection{The \boldmath{$\rho_3(1690)$}}

\begin{sloppypar}
The spin-3 resonance $\rho_3(1690)$ is described in terms of a totally symmetric
third-rank tensor field $\bm\rho_{\mu\nu\lambda} = \rho_{\mu\nu\lambda}^a \tau^a$ 
(of isospin $I=1$), which is subject to the constraints
\beq
\partial^\mu \bm\rho_{\mu\nu\lambda} = 0 ~, \qquad
g^{\mu\nu} \bm\rho_{\mu\nu\lambda} = 0 ~.
\eeq
The numerator of its propagator in momentum space involves the polarization sum~\cite{ZhuYan}
\begin{align}
P_{\mu_1\mu_2\mu_3}^{\nu_1\nu_2\nu_3} &= \frac{1}{6}\sum_{P\{\nu_1\nu_2\nu_3\}} \bigg\{
X_{\mu_1}^{\nu_1}X_{\mu_2}^{\nu_2}X_{\mu_3}^{\nu_3} 
- \frac{1}{5}\Big(
X_{\mu_1\mu_2}X^{\nu_1\nu_2}X_{\mu_3}^{\nu_3} \nnnl& \qquad\quad +
X_{\mu_1}^{\nu_1}X_{\mu_2\mu_3}X^{\nu_2\nu_3} +
X_{\mu_1\mu_3}X_{\mu_2}^{\nu_2}X^{\nu_1\nu_3} 
\Big)\bigg\} ~, \nnnl
X_{\alpha\beta} &= -g_{\alpha\beta} + \frac{p_\alpha p_\beta}{\mr^2} ~,
\end{align}
where the sum runs over all possible permutations of the indices $\{\nu_1,\nu_2,\nu_3\}$.
The simplest $\rho_3$ interaction Lagrangians 
with $\pi\pi$ and $\pi\omega$ are given by
\begin{align}
\Lagr_{\rho_3} &= \frac{g_{\rho_3\pi\pi}}{4\Fpi^2} \langle \bm\rho_{\mu\nu\lambda} 
\big[\partial^\mu\bm\pi,\partial^\nu\partial^\lambda\bm\pi\big] \rangle \nnnl
&+ \frac{g_{\rho_3\pi\omega}}{2\Fpi} \epsilon^{\lambda\alpha\beta\gamma}
\langle \bm\rho_{\mu\nu\lambda} \partial^\mu \partial_a \bm\pi \rangle 
\partial^\nu \partial_\beta \omega_\gamma ~, \label{eq:Lagr}
\end{align}
where $\bm\pi = \pi^a\tau^a$ denotes the isotriplet of pion fields, and $\omega_\mu$
the isosinglet $\omega$ vector field.
From Eq.~\eqref{eq:Lagr}, we can calculate the partial decay widths of the $\rho_3$ into 
$\pi\pi$ and $\pi\omega$ to be
\begin{align}
\Gamma(\rho_3 \to \pi\pi) &= \frac{g_{\rho_3\pi\pi}^2}{4480 \pi \Fpi^4\mr^2}\big(\mr^2-4\mpc^2\big)^{7/2} \,, \nnnl
\Gamma(\rho_3 \to \pi\omega) &= \frac{g_{\rho_3\pi\omega}^2}{13440 \pi \Fpi^2 \mr^7}
\lambda\big(\mr^2,M_\omega^2,\mpc^2\big)^{7/2} \,,
\end{align}
which show the expected phase-space dependence for the decay of a spin-3 particle.
From $\mr = (1688.8\pm 2.1)$\,MeV, $\Gamma_{\rho_3} = (161\pm 10)$\,MeV, 
$\BR(\rho_3\to\pi\pi) = (23.6\pm 1.3)\%$, $\BR(\rho_3\to\pi\omega) = (16\pm 6)\%$, and $\Fpi=92.2$\,MeV~\cite{PDG},
we therefore deduce
\beq
|g_{\rho_3\pi\pi}| = 0.056 \pm 0.005  \,, \quad
|g_{\rho_3\pi\omega}| = (1.2 \pm 0.5)\,\text{GeV}^{-2} \,. 
\eeq
\end{sloppypar}

\begin{figure}
\centering
\includegraphics[width=0.55\linewidth]{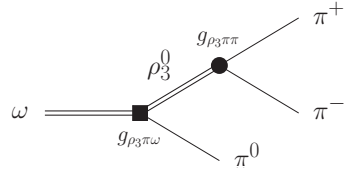}
\caption{Feynman diagram for the $\rho_3(1690)$-exchange contribution to $\omega\to 3\pi$ (in the $s$-channel).}
\label{fig:rho3}
\end{figure}
\begin{sloppypar}
Having fixed the coupling constants of the Lagran\-gian~\eqref{eq:Lagr}, we can proceed to calculate
the impact of the $\rho_3$ resonance on the decay $\omega \to 3\pi$.  
The exchange of a $\rho_3$ in the $s$-channel, see Fig.~\ref{fig:rho3}, yields a contribution 
to $\G(s)$ as defined in Eq.~\eqref{eq:defP+Famp} according to
\beq
\G_{\rho_3}(s) = C_F\,
\frac{\mr^2}{\mr^2-s} \frac{\kappa^2(s)}{M_\omega^4} ~, \quad
C_F = \frac{g_{\rho_3\pi\pi}g_{\rho_3\pi\omega}M_\omega^4}{60\Fpi^3 \mr^2} ~. \label{eq:Grho3}
\eeq
$\kappa^2(s)/M_\omega^4$ is a dimensionless kinematic factor characteristic for the F~wave.
The remaining effective coupling $C_F$ is numerically evaluated from the above to be
\beq
|C_F| \approx (1.4\pm 0.5)\times 10^{-4}\,\Fpi^{-3} ~.
\label{eq:Fcoupling}
\eeq
For illustration, we compare this to the simplified P-wave amplitude $\F(s)$ as given
by vector-meson dominance, e.g.\ according to the hidden-local-symmetry (HLS) 
formalism~\cite{Bando,Harada},
\beq
\F_{\rm HLS}(s) = C_P \,
\frac{M_\rho^2}{M_\rho^2-s} ~, \quad
C_P = \frac{N_c \, g}{8\pi^2\Fpi^3}  ~, \label{eq:HLS}
\eeq
where $N_c=3$ is the number of colors, $g \approx 5.8$ is the universal vector coupling,
and we have used the simplest choice for the anomalous HLS couplings, $c_1-c_2=c_3=1$~\cite{Harada,Benayoun1,Benayoun2}.
In this case, the effective coupling constant is
\beq
C_P \approx 0.22\,\Fpi^{-3} ~.\label{eq:Pcoupling}
\eeq
Comparing Eqs.~\eqref{eq:Fcoupling} and \eqref{eq:Pcoupling}, we see that the coupling constant alone,
i.e.\ before application of the F-wave phase-space factors, suppresses the F-wave vs.\ the P-wave
contribution by three orders of magnitude.
\end{sloppypar}

\subsection{The \boldmath{$\omega\to 3\pi$} F~wave}

\begin{figure}
\centering
\psfrag{Ghat}[c]{$C_P^{-1} M_\omega^4\kappa^{-2}(s)\times\hat\G_{\rm HLS}(s)$}
\psfrag{sqrt(s) [GeV]}{$\sqrt{s}$ [GeV]}
\includegraphics[width = 0.9\linewidth]{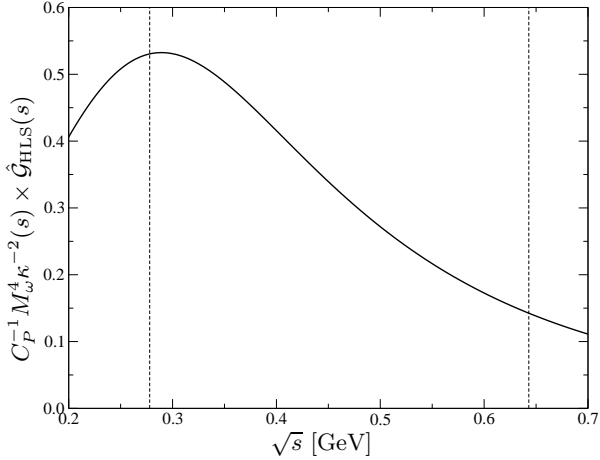}
\caption{The function $C_P^{-1} M_\omega^4\kappa^{-2}(s)\times\hat\G_{\rm HLS}(s)$, 
characterizing the F-wave projection of the $t$- and $u$-channel P-wave amplitudes; for details, see text.
The vertical dashed lines indicate the boundaries of the physical region for $\omega\to 3\pi$.}
\label{fig:Ghat}
\end{figure}

We therefore have strong indication that the $\omega\to 3\pi$ F~wave $f_3(s) = \G(s)+\hat\G(s)$
is dominated by the term in $\hat\G(s)$ given by the projection of the crossed-channel P-wave terms.
With the simplified amplitude~\eqref{eq:HLS}, we can calculate this contribution even analytically,
eschewing all complications of path deformation in the complete description.  The result is
\begin{align}
\hat\G_{\rm HLS}(s) &= -C_P \frac{2M_\rho^2}{M_\rho^2-\frac{1}{2}(3s_0-s)} \nnnl
&\times \frac{7}{8\bar\kappa^4}\bigg(\frac{5-6\bar\kappa^2+\bar\kappa^4}{2\bar\kappa}
\log\frac{1+\bar\kappa}{1-\bar\kappa} - \frac{15-13\bar\kappa^2}{3} \bigg) \,, \nnnl
\bar\kappa &= \frac{\kappa(s)}{2M_\rho^2-3s_0+s} ~.
\end{align}
In Fig.~\ref{fig:Ghat}, we plot the function $C_P^{-1} M_\omega^4\kappa^{-2}(s)\times\hat\G_{\rm HLS}(s)$,
which is seen to vary between 0.53 and 0.14 in the physical region $2\mpc \leq \sqrt{s} \leq M_\omega-\mpc$.
In view of Eq.~\eqref{eq:Grho3}, this is to be compared to the scale $C_F/C_P \approx 10^{-3}$
for the $\rho_3$-induced F-wave contribution.  
The conclusion is that we expect neglected terms to yield only percent-level corrections
even to the $\omega\to 3\pi$ F~wave.

It remains to ask to what extent the assumption of the F~wave being purely real is justified.
Ref.~\cite{Pelaez} provides a simple analytic parameterization of the $\pi\pi$ F-wave phase shift, 
according to which we find, at the upper limits of the phase space accessible in $\omega\to 3\pi$
and $\phi\to 3\pi$, 
\begin{align}
\delta_3^1\big(s=(M_\omega-\mpc)^2\big) &\approx 1\times 10^{-4} ~, \nnnl
\delta_3^1\big(s=(M_\phi-\mpc)^2\big) &\approx 7\times 10^{-4} ~,
\end{align}
which corresponds to $0.06^\circ$ and $0.4^\circ$, respectively.
The F-wave phase is tiny in all the accessible phase space; neglecting the imaginary part 
is an approximation way below the accuracy of our decay amplitude representation.

\subsection{Comparison to the pion--pion F~wave}

Finally, we wish to briefly point out that the hierarchy suggested for the $\omega\to 3\pi$
F~wave---complete dominance of $t$-channel exchange over the $s$-channel resonance tail
at low energies---should not come as a big surprise, looking at the comparable situation
in pion--pion scattering.  The $\rho_3$ contribution to the $\pi\pi$ F~wave that follows
from the Lagrangian~\eqref{eq:Lagr} is given by
\beq
t_3^1(s)_{\rho_3} = \frac{g_{\rho_3\pi\pi}^2}{8960\pi\Fpi^4}\frac{(s-4\mpc^2)^3}{\mr^2-s} ~.
\eeq
This yields a contribution to the F-wave scattering length according to
\beq
(a_3^1)_{\rho_3} = \frac{g_{\rho_3\pi\pi}^2}{140\pi\Fpi^4\mr^2}
\approx 2.6\times 10^{-7} \mpc^{-6} ~,
\eeq
to be compared with a phenomenological value of $a_3^1 \approx 5.4\times 10^{-5}\mpc^{-6}$~\cite{ACGL}:
the $\rho_3$ contribution to the scattering length is suppressed by more than two orders of magnitude.
The leading-order prediction in chiral perturbation theory at one loop, in contrast, yields~\cite{GL83}
\beq
a_3^1 = \frac{11}{94080\pi^3\Fpi^4\mpc^2} + \Order(\mpc^0) \approx 2.0\times 10^{-5}\mpc^{-6}~,
\eeq
which is entirely given by ($t$- and $u$-channel) loop effects.

\section{Properties of the dispersion integral}\label{app:angint}

\subsection{The angular integral}

The dispersion relation Eq.~\eqref{eq:inteeq} is constructed for the corresponding elastic scattering process $V\pi\to\pi\pi$ 
with $\mV<3\mpc$, exploiting unitarity.
We will now discuss the analytic continuation to the decay process and its impact on the way 
the angular integration in the calculation of $\hat\F(s)$ is performed.
Firstly, we note that we require the single-variable amplitude $\F(s)$ in the complex plane for the iteration procedure.
Secondly, we need to continue to the physical vector-meson masses. That means that we have to perform the following continuation:
\begin{equation}
\lim_{\mV \to \mV^{\mathrm{phys}}}\lim_{s\to\mathbb{C}}\F(s)~.
\end{equation}
Such a continuation was discussed in Ref.~\cite{Bronzan}, comparing the dispersive representation to perturbation theory. 
Summarizing this analysis, we have to give $\mV^2$ an infinitesimal positive imaginary part
($\mV^2\to\mV^2+i\epsilon$). 
Apart from the angular average integration Eq.~\eqref{eq:inhomogen}, the involved functions depend trivially on $\mV^2$ and hence 
the continuation of these functions is simple. The continuation of the angular integral on the other hand is non-trivial.

The first issue can be deduced directly from the defining equation of the angular average integrals Eq.~\eqref{eq:inhomogen}. 
If the argument of the integrand has a real part greater than
$4 \mpc^2$, the integration contour may cross the right-hand cut of $\F(s)$, necessitating its deformation. 
This issue arises in the continuation to vector-meson masses greater than $3\mpc$ as a consequence of the three-particle cut,
and is therefore an inherent property of a \emph{decay} (as opposed to scattering) amplitude.
For closer inspection we rewrite Eq.~\eqref{eq:inhomogen} in the following form:
\begin{equation}\label{eq:angint}
\langle z^{n}\F(s) \rangle= \frac{1}{\kappa(s)}\int_{s_-(s)}^{s_+(s)} ds' \left(\frac{2s'-3s_{0}+s}{\kappa(s)}\right)^{n} \F(s')~,
\end{equation}
with the integration limits according to the convention $\mV^2\to\mV^2+i\epsilon$
\begin{equation}
 s_{\pm}(s,\mV^2+i\epsilon)=s_{\pm}(s,\mV^2)+i\epsilon\frac{\partial s_{\pm}(s,\mV^2)}{\partial \mV^2}~,
\end{equation}
we have (see e.g.\ Ref.~\cite{Kambor})
\begin{align}
2s_{+}(s)&=\left\{\begin{array}{ll} 3s_{0}-s+|\kappa(s)|+i\epsilon\,, &~ s\in[4\mpc^2,a] \,,\\[2mm] 
3s_{0}-s+i|\kappa(s)| \,,&~ s\in[a,b] \,,\\[2mm]
3s_{0}-s-|\kappa(s)| \,,&~ s\in[b,\infty] \,, \end{array}\right.\notag \\
2s_{-}(s)&=\left\{\begin{array}{ll} 3s_{0}-s-|\kappa(s)|+i\epsilon \,,&~ s\in\Big[4\mpc^2,\frac{\mV^2-\mpc^2}{2}\Big] \,,\\[2mm]
3s_{0}-s-|\kappa(s)|-i\epsilon \,,&~ s\in\Big[\frac{\mV^2-\mpc^2}{2},a\Big] \,, \\[2mm]
3s_{0}-s-i|\kappa(s)| \,,&~ s\in[a,b] \,, \\[2mm]
3s_{0}-s+|\kappa(s)| \,,&~ s\in[b,\infty] \,, \end{array}\right.
\end{align}
where
\beq
a\equiv(\mV-\mpc)^2 ~, \quad  b\equiv(\mV+\mpc)^2 ~.
\eeq
Care has to be taken at the singularities of the Jacobian ($\kappa(s)=0$), as here the above parameterization breaks down.

\begin{figure}
\psfrag{Re}[tl][tl][1.1]{\hspace*{-2mm}$\Re s_\pm(s)$}
\psfrag{Im}[bl][l][1.1]{$\Im s_\pm(s)$}
\includegraphics[width = 0.95\linewidth]{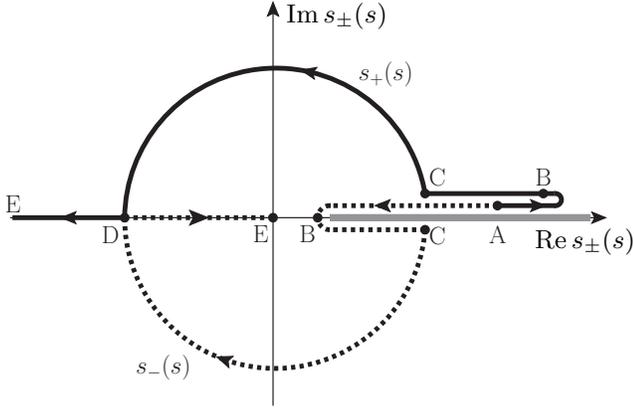}
\caption{Trajectories of the integration limits in Eq.~\eqref{eq:angint}; see text for details.}
\label{fig:Endpoint}
\end{figure}
We now deduce where an integration path deformation is needed to avoid crossing the right-hand cut.
The trajectories of the integration limits in the complex plane are depicted in Fig.~\ref{fig:Endpoint}.
In the scattering region $s\in[b,\infty]$ (corresponding to $[\text{D},\text{E}]$ in Fig.~\ref{fig:Endpoint}) we have the standard partial-wave projection with line integrations. 
In the unphysical region $s\in[a,b]$ ($[\text{C},\text{D}]$) we have to deform the integration contour in the 
vicinity of $a$ (C), as $\Re s_+(s)$, $\Re s_-(s) >4\mpc^2$ and $\Im s_+(s) >0>\Im s_-(s)$.
Since we have the trajectories of the integration limits at hand, 
we can use these as parameterizations for the integration contour  
and exploit the fact that $s_+$ and $s_-$ meet at D. 
We parameterize the integration from $s_-(s)$ to $s_+(s)$ as follows: starting from the point $s_-(s)$, we follow the  $s_-$ trajectory up to D, then use the $s_+$ trajectory to $s_+(s)$.
In the interval $s\in[(\mV^2-\mpc^2)/2,a]$ ($[\text{B},\text{C}]$), $s_-(s)$ is below the cut, while $s_+(s)$ is above. In this part of the decay region
we deform the integration contour as follows: the contour follows $s_-$ up to B, then bypasses the branch point $4\mpc^2$ and moves to the upper integration limit above the cut via $s_+$. 
In the interval $s\in[4\mpc^2,(\mV^2-\mpc^2)/2]$ ($[\text{A},\text{B}]$) we can integrate via lines, as the upper and lower integration limits lie on the upper rim of the cut.

\subsection{Analytic structure of the integrand}\label{app:struct}

In this section we wish to briefly discuss the analytic structure of the integrand of the dispersion integral in Eq.~\eqref{eq:inteeq}, in particular what effects are generated by the fact that the vector
particle can decay into the final-state particles. The analytic structure of the integrand is fully determined by the properties of the function $\kappa(s)$ and its zeros. The latter are the two-particle 
scattering thresholds $4\mpc^2$ and $(\mV+\mpc)^2$, and the pseudo-threshold $(\mV-\mpc)^2$. 

\begin{figure}[!t]
 \centering 
 \psfrag{sqrt(s) [GeV]}[c][b][1]{$\sqrt{s}~{\rm[GeV]}$}
 \includegraphics[width= 0.98\linewidth]{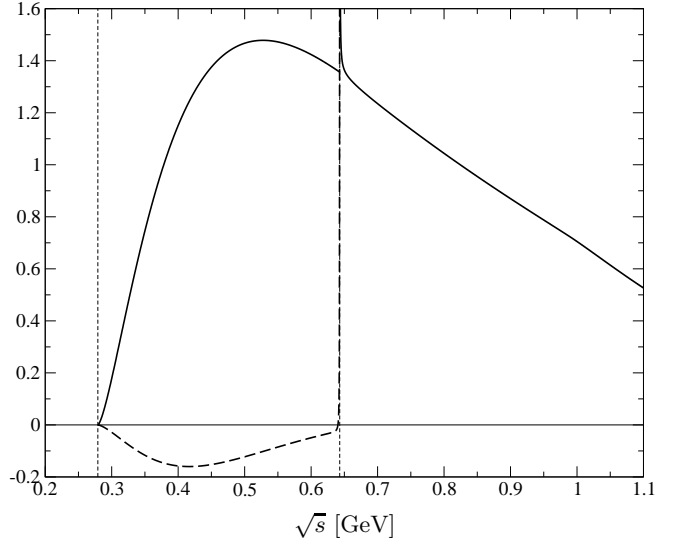}
 \caption{Real (solid line) and imaginary (dashed line) part of the integrand as a function of $\sqrt{s}$. The occurrence of the singularities at the pseudo-threshold (dashed vertical line) is explained in 
the text.}
\label{fig:singularity}
\end{figure}

Due to the particular structure of the angular integral in Eq.~\eqref{eq:angint} the zeros of $\kappa(s)$ in principle give rise to singularities of square-root order. However, the physical thresholds at 
$s=4\mpc^2$ and $s=(\mV+\mpc)^2$ do not contribute to the angular integral, since the integration path is shrunk to a point. This is not the case at the pseudo-threshold. Due to the cut generated by
the two-pion threshold and the resulting path deformation the integral gives rise to a non-vanishing contribution and thus to the aforementioned singularity. As explained in the previous section the path
deformation adheres to the ability of the vector particle to decay into the final-state particles, that is, $\mV>3\mpc$. In Fig.~\ref{fig:singularity} we display the integrand of the dispersion integral. 

Notice that our observation is in complete agreement with a study on the properties of the non-trivial relativistic two-loop graph given in Ref.~\cite{GKR}. There the singularity is explained
with a deformation of the path of the angular integration to infinity as $s$ approaches the pseudo-threshold. This is analogous to the previous argument.

\section{Contribution from higher resonances}\label{app:highres}

In order to estimate the potential influence of higher P-wave resonances 
($\rho'(1450)$, $\rho''(1700)$) on the $V \to 3\pi$ decays, we follow 
an approach recently suggested in connection with the pion vector form factor~\cite{Roig}.
There, the following form factor representation has been fitted to the 
high-precision data for $\tau^- \to \pi^-\pi^0\nu_\tau$ from Belle~\cite{BelleFF}:
\begin{align}
&F_\pi^V(s) 
= \frac{M_\rho^2+s(\gamma\, e^{i\phi_1}+\delta\, e^{i\phi_2})}{M_\rho^2-s-iM_\rho\Gamma_\rho(s)} 
\exp\bigg\{\!-\frac{s A_\pi(s)}{96\pi^2\Fpi^2}\bigg\} \nnnl
& - \frac{\gamma\, s\, e^{i\phi_1}}{\mrr^2-s-i\mrr\grr(s)} 
\exp\bigg\{\!-\frac{s\grr A_\pi(s)}{\pi \mrr^3\sigma_\pi^3(\mrr^2)}\bigg\} \nnnl
& - \frac{\delta\, s\, e^{i\phi_2}}{\mrrr^2-s-i\mrrr\grrr(s)} 
\exp\bigg\{\!-\frac{s\grrr A_\pi(s)}{\pi \mrrr^3\sigma_\pi^3(\mrrr^2)}\bigg\} \,,
\label{eq:roigFF}
\end{align}
where
\begin{align}
A_\pi(s) &= \log\frac{M_\pi^2}{M_\rho^2} + \frac{8M_\pi^2}{s} - \frac{5}{3} 
+ \sigma_\pi^3(s)\log\frac{1+\sigma_\pi(s)}{1-\sigma_\pi(s)} \,, \nnnl
\Gamma_\rho(s) &= \frac{M_\rho s}{96\pi\Fpi^2} \sigma_\pi^3(s) \,, \nnnl
\Gamma_{\rho',\rho''}(s) & = \frac{M_{\rho',\rho''}}{\sqrt{s}}
\bigg(\frac{s-4\mpc^2}{M_{\rho',\rho''}^2-4\mpc^2}\bigg)^{3/2}
\Gamma_{\rho',\rho''} \,.
\end{align}
We have omitted a kaon-loop contribution to $\Gamma_\rho(s)$
(which also affects the real exponential multiplying the $\rho$ propagator in 
Eq.~\eqref{eq:roigFF}) that is retained in Ref.~\cite{Roig}:
it does not have a large effect on the form factor, and 
we want to translate the above representation into a single-channel
Omn\`es form with only elastic $\pi\pi$ rescattering included; 
Eq.~\eqref{eq:roigFF} treats the $\rho'$ and $\rho''$ as purely elastic resonances
anyway.  Besides, inelasticities in the $\pi\pi$ P~wave in general, and in these higher
P-wave resonances in particular, are expected to be dominated by $4\pi$
rather than $K\bar K$.  Furthermore, our form of Eq.~\eqref{eq:roigFF} is only valid
above threshold, $s \geq 4\mpc^2$.

\begin{figure}
\centering
\psfrag{delta(s)}{$\delta_1^1(s)$}
\psfrag{sqrt(s) [GeV]}{$\sqrt{s}$ [GeV]}
\includegraphics[width = 0.94\linewidth]{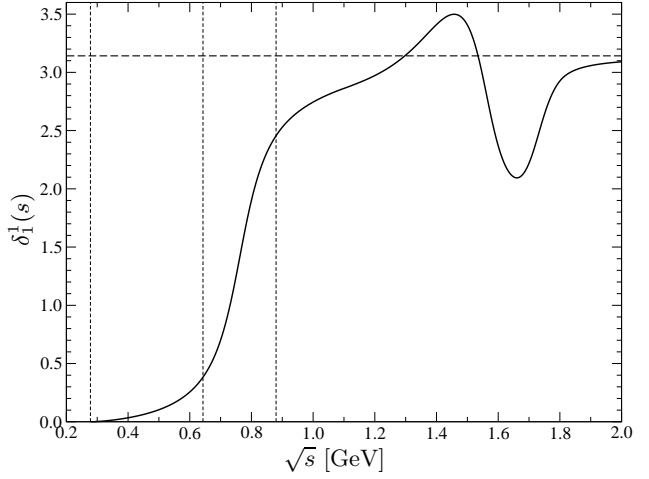}
\caption{Alternative P-wave phase shift $\delta_1^1(s)$ including the effects of $\rho'(1450)$ 
and $\rho''(1700)$ resonances; see text for details.  The horizontal dashed line denotes
the assumed asymptotic phase of $\pi$, the vertical dashed lines show the boundaries of the
physical region for the two decays at $\sqrt{s}=2\mpc$, $M_\omega-\mpc$, $M_\phi-\mpc$.}
\label{fig:phase_rhoprime}
\end{figure}

Given the treatment of $\rho'$, $\rho''$ as \emph{elastic} resonances, 
we implement these into our formalism by just modifying the $\pi\pi$ P-wave phase shift
accordingly and using the phase of the form factor in Eq.~\eqref{eq:roigFF} as the input
for a modified Omn\`es function.  
More precisely, we employ a phase identical to the one of Ref.~\cite{CapriniWIP} 
below $\sqrt{s}=1.1$\,GeV, guide this smoothly to the phase of the form factor,
which is used up to $\sqrt{s}=1.8$\,GeV (roughly the kinematic range accessible in
$\tau^- \to \pi^-\pi^0\nu_\tau$ and therefore fitted in Ref.~\cite{Roig}), before
guiding the phase smoothly to $\pi$. This phase, calculated for
the central parameter values of the fit to the Belle data~\cite{BelleFF}
given by
$\mrr =1.461$\,GeV, 
$\grr =0.353$\,GeV, 
$\mrrr=1.732$\,GeV, 
$\grrr=0.141$\,GeV, 
$\gamma = 0.088$, $\delta = -0.024$,
$\phi_1 = 0.6$, $\phi_2 = 0.8$~\cite{RoigPC},
is shown in Fig.~\ref{fig:phase_rhoprime}.

We refrain from investigating the variation within the error ranges 
on these parameters determined in Ref.~\cite{Roig}: 
we regard our estimate of higher-resonance contributions as indicative rather
(and, as such, they will turn out to be entirely negligible) than as a 
precision study; which, as we will argue below, it cannot be as a matter of principle.
We would like to point out  that the form factor representation~\eqref{eq:roigFF}
allows to lead the phase smoothly to $\pi$: the higher resonances only induce ``wiggles''
in the phase.  This is in contrast to the (weighted) sum of Breit--Wigner or
Gounaris--Sakurai~\cite{GounarisSakurai} functions as employed in the data fit
in Ref.~\cite{BelleFF}, which makes the phase rise by $\pi$ for each resonance, leading
to an asymptotic phase of rather $3\pi$ in the case at hand.

It turns out that the effect of using an alternative Omn\`es function, based on the phase
shown in Fig.~\ref{fig:phase_rhoprime}, as the starting point for our
dispersive representation of the $V \to 3\pi$ Dalitz plots is totally negligible:
it changes the Dalitz plot distributions by less than 1\%, and is therefore smaller
than the other uncertainty sources identified in the main text.
Figure~\ref{fig:phase_rhoprime} makes this rather plausible: the deviations due to the two
high-mass resonances show up only significantly above the energy range accessible
in the $\omega$ and $\phi$ decays; furthermore, in the integral, the two ``wiggles'' of opposite
sign tend to cancel out.  Indeed, calculating e.g.\ the sum-rule value for the pion charge
radius according to Eq.~\eqref{eq:sumrule}, we find
\beq
\langle r_{\text{sum}(\rho',\rho'')}^2 \rangle^V_\pi\simeq 0.420\ {\text{fm}}^2~,
\eeq
very close to the value $0.415\ {\text{fm}}^2$ found with our ``standard phase''.

We need to recall the rather indicative nature of this investigation. 
We employ information on the $\rho'$ and $\rho''$ resonances gained experimentally
from the pion form factor measurement in $\tau^- \to \pi^- \pi^0 \nu_\tau$~\cite{BelleFF};
it might be possible that they couple particularly strongly to $\omega\pi$, and thus
have a more pronounced impact on $\omega\to 3\pi$.  However, phenomenological information
does not seem to support this: if we compare the pion form factor fit of Ref.~\cite{BelleFF}
with an experimental analysis of $e^+e^-\to\omega\pi^0$~\cite{CMD2-omegapi}, both 
employing weighted sums of Breit--Wigner or Gounaris--Sakurai resonance propagators as fit functions, 
the pion form factor suggests a (modulus of the) coupling strength of the $\rho'$ relative
to the $\rho$ of the order of 15\%, while the $\omega\pi$ production measurement
suggests a $\rho'$-to-$\rho$ ratio of the order of 10\% (despite the fact that here, 
the $\rho$ is obviously sub-threshold, and phase space tends to emphasize the higher-resonance
contributions in the cross section).

We therefore conclude that the influence of higher-mass P-wave resonance states on the $V\to3\pi$ 
decays seems to be very small, and well contained in the uncertainty estimate we have performed
in the main text.

\section{Inelasticity parameter}\label{app:inpara}

In this appendix, we wish to briefly derive Eq.~\eqref{eq:inteeqinhom}, following Ref.~\cite{Walker}.
We introduce an inelasticity parameter \(\eta(s)\) in the parameterization of the $\pi\pi$ 
P-wave amplitude (see Eq.~\eqref{eq:inelpara}) according to
\begin{equation}
 \sin\delta(s) e^{-i\delta(s)} \rightarrow \frac{1}{2i}\bigl( 1-\eta(s)e^{-2i\delta(s)} \bigr)~.
\end{equation}
Thus the unitarity relation for $\F(s)$ takes the following modified form
\begin{equation}\label{eq:modunitrel}
 \disc\F(s) = \bigl(\F(s)+\hat\F(s)\bigr)\theta(s-4\mpc^2)\Big(1-\eta(s)e^{-2i\delta(s)}\Big)~.\\
\end{equation}
For the homogeneous case \(\hat\F(s)=0\), we take the logarithm
\begin{equation}
 \disc \log\F(s) = 2i\,\delta(s)-\log\eta(s)~,
\end{equation}
which leads to the  modified Omn\`es solution
\begin{align}
\tilde\Omega(s)&=\Omega(s)\exp\biggl\{\frac{is}{2\pi}\int\limits_{16\mpc^2}^\infty ds'\frac{\log\eta(s')}{s'(s'-s-i\epsilon)}\biggr\}\notag\\
	       &=\xi(s)\Xi(s)\Omega(s)~,
\end{align}
with 
\begin{align}
\xi(s)= \begin{cases}
 \eta^{-1/2}(s) &\text{above the cut,}\\
 \eta^{1/2}(s) &\text{below the cut,}\\
 1 &\text{elsewhere,}
\end{cases}
\end{align}
and
\begin{equation}
 \Xi(s)=\exp\biggl\{\frac{is}{2\pi}\dashint{10pt}\limits_{16\mpc^2}^\infty ds'\frac{\log\eta(s')}{s'(s'-s)}\biggr\}~.
\end{equation}
For the full solution of the unitarity relation~\eqref{eq:modunitrel} we use the product ansatz \(\F(s)=\tilde\Omega(s)\psi(s)\)
to obtain
\begin{equation}
\disc \psi(s)=\frac{\hat\F(s)\bigl(e^{i\delta(s)}-\eta(s)e^{-i\delta(s)}\bigr)}{\sqrt{\eta(s)}\Xi(s)|\Omega(s)|}~.
\end{equation}
Rewriting \(\psi(s)\) into a dispersion relation finally leads to the full solution quoted in Eq.~\eqref{eq:inteeqinhom}.

\end{document}